\documentclass[a4paper,11pt]{article}
\pdfoutput=1 

\usepackage{jcappub} 

\usepackage[T1]{fontenc} 

\usepackage{graphicx}
\usepackage{dcolumn}
\usepackage{bm}
\usepackage{physics}
\usepackage{xcolor}
\usepackage{comment}
\usepackage{amsmath}

\newcommand{\mpl}{M_{\mathrm{pl}}}
\newcommand{\bx}{\mathbf{x}}
\newcommand{\bp}{\mathbf{p}}
\newcommand{\bq}{\mathbf{q}}
\newcommand{\bk}{\mathbf{k}}
\newcommand{\md}{\mathrm{d}}

\title{\boldmath Perturbative region on non-Gaussian parameter space in single-field inflation}

\author[1,2]{Jason Kristiano}
\author[1,2,3,4]{and Jun'ichi Yokoyama}

\affiliation[1]{Research Center for the Early Universe (RESCEU), Graduate School of Science, The University of Tokyo, Tokyo 113-0033, Japan}
\affiliation[2]{Department of Physics, Graduate School of Science, The University of Tokyo, Tokyo 113-0033, Japan}
\affiliation[3]{Kavli Institute for the Physics and Mathematics of the Universe (Kavli IPMU), WPI, UTIAS, The University of Tokyo, Kashiwa, Chiba 277-8568, Japan}
\affiliation[4]{Trans-Scale Quantum Science Institute, The University of Tokyo, Tokyo 113-0033, Japan}

\emailAdd{jkristiano@resceu.s.u-tokyo.ac.jp}
\emailAdd{yokoyama@resceu.s.u-tokyo.ac.jp}

\abstract{
We calculate one-loop correction to the two-point functions of curvature perturbation in single-field inflation generated by cubic self-interaction. Incorporating the observed red-tilted spectrum of curvature perturbation, the relevant one-loop correction takes a finite value and inversely proportional to the spectral tilt. Requiring one-loop correction to be much smaller than the tree-level contribution leads to an upper bound on primordial non-Gaussianity. While observationally allowed region of non-Gaussian parameter space is found to be entirely included by the region, where one-loop correction is smaller than the tree-level contribution, an appreciably large region has one-loop correction larger than 1\% or even 10\% of the latter. If future observations conclude non-Gaussianity falls in such a region, then it would be important to incorporate higher-order  corrections to the spectrum in order to achieve  precise cosmology.  In some extreme cases, where one-loop correction has a comparable magnitude to the tree-level contribution, it might indicate breakdown of the cosmological perturbation theory in the context of single-field inflation.
}

\begin{document}
\maketitle
\flushbottom

\section{Introduction}

Observation of cosmic microwave background (CMB) shows that its temperature is very isotropic with small fluctuations of order of $10^{-5}$ times the background temperature \cite{Akrami:2018vks,Akrami:2018odb}. The homogeneity of CMB can be explained if our Universe underwent a very rapid accelerated expansion before the Big Bang, known as inflation \cite{Starobinsky:1980te, Sato:1980yn, Guth:1980zm}. At the same time, small fluctuations are generated from quantum fluctuations of spacetime and stretched by inflation \cite{Starobinsky:1979ty,Mukhanov:1981xt,Starobinsky:1982ee,Hawking:1982cz,Guth:1982ec}. This framework is called cosmological perturbation theory. Observation also shows the power spectrum of fluctuations is almost scale-invariant with a highly Gaussian distribution. Deviation from scale-invariant power spectrum is called spectral-tilt and has been observed fairly well. Deviation from Gaussian distribution is called primordial non-Gaussianity, and so far only upper-bound is obtained observationally \cite{Akrami:2019izv}.

Theoretically, such quantum fluctuations are analyzed within the framework of quantum field theory (QFT). Power spectrum corresponds to the vacuum expectation value (VEV) of two-point functions of the fluctuations. At the lowest order, primordial non-Gaussianity corresponds to the VEV of three-point functions of fluctuations. This originates from cubic self-interactions of the fluctuations. At the same time, such self-interaction can generate backreaction to the two-point functions. In the terminology of QFT, the backreaction is called loop correction. 

We are interested in models predicting a sizable amount of non-Gaussianity close to the observational bound \cite{Akrami:2019izv}, as such models may induce large loop corrections of inflationary power spectrum.  The simplest inflation model, namely canonical slow-roll inflation, predicts a small non-Gaussianity \cite{Maldacena:2002vr}. Its bispectrum is proportional to the slow-roll parameters. Therefore, we have to look for other inflation models. There are some inflation models which can realize a large non-Gaussianity by modifying kinetic terms, such as $k$- and $G$-inflation \cite{ArmendarizPicon:1999rj,Garriga:1999vw,Kobayashi:2010cm,Kobayashi:2011nu}, ghost condensate \cite{ArkaniHamed:2003uy,ArkaniHamed:2003uz}, and Dirac-Born-Infeld inflation \cite{Alishahiha:2004eh}. For inflation models that generate large loop correction, we must require that the one-loop correction is small enough compared to the tree-level two-point functions, otherwise we cannot trust the perturbation theory because the power spectrum would be infinitely large if we calculated higher-order loop correction.

Weinberg first realized the case that loop correction can become significantly large \cite{Weinberg:2005vy, Weinberg:2006ac}. He calculated one-loop correction generated by coupling between curvature perturbations and massless scalar fields. For perturbation with comoving wavenumber $k$, he found the one-loop correction proportional to $\log k$. Another notable calculation was performed by Senatore and Zaldarriaga \cite{Senatore:2009cf}, where they calculated one-loop correction generated by a cubic self-interaction of curvature perturbations. Assuming a scale-invariant curvature perturbation, they performed the calculation by two different regularization methods, physical cutoff and dimensional regularization. They found that both methods lead to the same one-loop correction results proportional to $\log H$, where $H$ is Hubble scale during inflation. After that, full calculation of one-loop correction generated by all possible self-interaction of curvature perturbations was done by Bartolo, et. al. \cite{Bartolo:2010bu}, where they relied on the regularization method by Senatore and Zaldarriaga. Another way to obtain one-loop correction is by bootstrap method as demonstrated by \cite{Melville:2021lst}, relying on the cosmological optical theorem \cite{Goodhew:2020hob}.

In our previous letter \cite{Kristiano:2021urj}, we calculated one-loop correction generated by a cubic self-interaction of curvature perturbations. By properly incorporating the observed red tilt of the spectrum of curvature perturbation, we found that the cosmologically relevant one-loop correction has a finite contribution inversely proportional to the spectral-tilt. Note that Senatore and Zaldarriaga \cite{Senatore:2009cf} assumed the scale-invariant curvature perturbation, and consequently they found a divergent one-loop correction. They performed regularization and renormalization to remove the divergence and extract the finite part of it. In our approach, we do not perform regularization and renormalization because the relevant one-loop correction is finite. Then, we obtained an upper bound on primordial non-Gaussianity by requiring one-loop correction to be much smaller than the tree-level two-point functions. 

In this paper, we perform the full calculation of one-loop correction generated by cubic self-interaction of curvature perturbations from the effective field theory of inflation \cite{Cheung:2007st}. In section 2, we review the standard cosmological perturbation theory. In section 3, we calculate the one-loop correction to the power spectrum. In section 4, we plot the perturbative region in the non-Gaussian parameter space and compare it with observational constraint. Finally in section 5, we conclude our paper. We show the detailed calculation in the appendix.

\section{Cosmological Perturbation Theory}

In this section, we briefly review the standard cosmological perturbation theory. We consider the simplest single-field inflation model that can generate a large non-Gaussianity, namely  $k$ inflation or $P(X,\phi)$ inflation. It is an inflation model whose Lagrangian is a general function of $\phi$ and $X \equiv - \nabla_\mu \phi \nabla^\mu \phi/2$, with the action  
\begin{equation}
S = \frac{1}{2} \int \mathrm{d}^4x \sqrt{-g} \left[ M_{\mathrm{pl}}^2 R + 2P(X,\phi) \right],
\end{equation}
where $g = \mathrm{det} (g_{\mu \nu})$, $g_{\mu \nu}$ is a metric tensor, and $R$ and $M_{\mathrm{pl}}$ are the Ricci scalar and the reduced Planck scale, respectively. The Einstein equations in the homogeneous and isotropic background with the metric $\md s^2 = -\md t^2 + a^2(t) \md \bx^2$ read
\begin{equation}
H^2 = \frac{1}{3} \left( 2 \bar{X} \bar{P}_{,X} - \bar{P} \right) ~~\mathrm{and}~~ \dot{H} = -\bar{X} \bar{P}_{,X}, \label{friedmann3}
\end{equation}
where $H = \dot{a}/a$, a dot denotes time derivative, a comma represents a partial derivative, and $\bar{X} = \dot{\bar{\phi}}^2/2$ is the kinetic energy of the background scalar field with $\bar{P} = P(\bar{X}, \bar{\phi})$.

\subsection{Second-order Action}
We now incorporate small perturbation to the homogeneous and isotropic background,
expressing the scalar field as $\phi(\mathbf{x},t) = \bar{\phi}(t) + \delta \phi(\mathbf{x},t)$.
It is convenient to use the Arnowitt-Deser-Misner (ADM) formalism where the perturbed metric is 
expressed as
\begin{equation}
ds^2 = -N^2 \mathrm{d}t^2 + \gamma_{ij} (\mathrm{d}x^i + N^i \mathrm{d}t)(\mathrm{d}x^j + N^j \mathrm{d}t).
\end{equation}
Here $\gamma_{ij}$ is a three-dimensional metric on slices of constant $t$, $N$ is a lapse function, and $N^i$ is a shift vector. Then, the action becomes
\begin{equation}
\label{admaction}
S = \frac{1}{2} \int \mathrm{d}^4x \sqrt{\gamma} \left[ N {}^{(3)}R + 2 N P(X,\phi) + N^{-1} (E_{ij}E^{ij} - E^2) \right],
\end{equation}
where $\gamma = \mathrm{det} ~\gamma_{ij}$, ${}^{(3)}R$ is Ricci scalar of $\gamma_{ij}$,
\begin{equation}
E_{ij} = {\textstyle \frac{1}{2}} (\dot{\gamma}_{ij} - \nabla_i N_j - \nabla_j N_i), ~~\mathrm{and}~~ E = E^i_i.
\end{equation}
$\nabla_i$ is three-dimensional covariant derivative of the metric $\gamma_{ij}$. However, such perturbations are very general and contain gauge freedom. We can choose to work in the comoving gauge
\begin{equation}
\delta \phi(\mathbf{x},t) = 0, ~\gamma_{ij}(\mathbf{x},t) = a^2(t)[1 + 2\zeta(\mathbf{x},t)] \delta_{ij},
\end{equation}
where $\zeta(\mathbf{x},t)$ is the comoving curvature perturbation. Then, $N$ and $N^i$ are obtained by solving the Hamiltonian and momentum constraint equations
\begin{gather}
\nabla_i [N^{-1} (E^i_j - \delta^i_j E)] = 0, \nonumber\\
{}^{(3)}R + 2P - 4X P_{,X} - N^{-2} (E_{ij}E^{ij} - E^2) = 0,
\end{gather}
respectively. $N_i$ is decomposed into two parts, $N_i = \tilde{N}_i + \partial_i \psi$ where $\partial_i \tilde{N}^i = 0$. Then, $N$ and $N^i$ are expanded in terms of powers of $\zeta$ as
\begin{gather}
N = 1 + \alpha_1 + \alpha_2 + \dots \nonumber\\
\tilde{N}_i = \tilde{N}_i^{(1)} + \tilde{N}_i^{(2)} + \dots \nonumber\\
\psi = \psi_1 + \psi_2 + \dots \label{nexpand}
\end{gather}
where $\alpha_n, \tilde{N}^{(n)}, \psi_n \sim O(\zeta^n)$. The first-order solution of the constraint equations are
\begin{equation}
\alpha_1 = \frac{\dot{\zeta}}{H}, ~~\tilde{N}_i^{(1)} = 0, ~~\mathrm{and}~~ \psi_1 = - \frac{\zeta}{H} + a^2 \frac{\epsilon}{c_s^2} \partial^{-2} \dot{\zeta},
\end{equation}
where inverse Laplacian $\partial^{-2}$ is defined as $\partial^{-2}(\partial^2 \zeta) = \zeta$ and $c_s$ is the sound speed given by
\begin{equation}
c_s^2 = \frac{\bar{P}_{,X}}{\bar{P}_{,X} + 2 \bar{X} \bar{P}_{,XX}}.
\end{equation}
Here $\epsilon\equiv -\dot{H}/{H^2}$ is a  slow-roll parameter of the Hubble parameter $H$,
which has a very weak time dependence during inflation. Later, we will also use another slow-roll parameter $\eta = \dot{\epsilon}/\epsilon H$, which represents the change of slow-roll parameter $\epsilon$ during inflation. After substituting the first-order solution for $N$ and $N^i$ to the action and performing cumbersome partial integration, we obtain the following second-order action of the curvature perturbation
\begin{equation}
S^{(2)} = M_{\mathrm{pl}}^2 \int \mathrm{d}t ~\mathrm{d}^3x ~a^3 \frac{\epsilon}{c_s^2} \left[ \dot{\zeta}^2 - \frac{c_s^2}{a^2} (\partial_i \zeta)^2  \right].
\label{S2}
\end{equation}

By introducing the Mukhanov-Sasaki (MS) variable \cite{Mukhanov:1981xt,Sasaki:1986hm}
\begin{equation}
v \equiv M_{\mathrm{pl}}z\zeta , ~~~~z \equiv \frac{a}{c_s}\sqrt{2 \epsilon},
\end{equation}
the second-order action (\ref{S2}) becomes canonically normalized
\begin{equation}
S^{(2)} = \frac{1}{2} \int \mathrm{d}\tau ~\mathrm{d}^3x \left[ (v')^2 - c_s^2 (\partial_i v)^2 + \frac{z''}{z} v^2 \right],
\end{equation}
where $\md \tau = a ~\md t$, $\tau$ is conformal time, and the prime denotes derivative with respect to $\tau$. The classical equation of motion of $v(\bx, \tau)$ is
\begin{equation}
v''(\bx, \tau) + \left( -c_s^2 \partial_i \partial^i + \frac{z''}{z} \right) v(\bx, \tau) = 0.
\end{equation}
Performing Fourier transformation
\begin{equation}
v(\bx, \tau) = \int \frac{\md^3 p}{(2 \pi)^3} v(\bp, \tau) e^{-i \bp \cdot \bx},
\end{equation}
the equation of motion in momentum space is
\begin{equation}
\label{MSeq}
v''(\bp, \tau) + \left( c_s^2 p^2 - \frac{z''}{z} \right) v(\bp, \tau) = 0,
\end{equation}
where $p$ is defined as $p = |\bp|$. Such equation is called MS equation. To solve it, we need the time dependence of $z$, which up to the lowest order in the slow-roll parameters is given by
\begin{equation}
\frac{z''}{z} = \frac{1}{\tau^2} \left( 2 + 3\epsilon + \frac{3\eta}{2} \right).
\end{equation}
Quantization is performed by promoting the MS variable to an operator
\begin{equation}
\hat{v}(\mathbf{p}, \tau) =M_{\mathrm{pl}}z\zeta(\mathbf{p}, \tau)=  v_p(\tau) \hat{a}_{\mathbf{p}} + v^*_p (\tau) \hat{a}_{-\mathbf{p}}^\dagger, \nonumber
\end{equation}
where $v_p(\tau)$ is the mode function, superscript $*$ denotes complex conjugate operation, $\hat{a}_{-\mathbf{p}}^\dagger$ and $\hat{a}_{\mathbf{p}}$ are creation and annihilation operators with the commutation relation
\begin{equation}
\left[ \hat{a}_{\mathbf{p}}, \hat{a}_{-\mathbf{q}}^\dagger \right] = (2 \pi)^3 \delta^3(\mathbf{p} + \mathbf{q}).
\end{equation}
The mode function is the solution of MS equation
\begin{equation}
v_p(\tau) = \left(-\frac{\pi\tau}{4}\right)^{1/2} \exp\left[ \frac{i\pi}{4} (2\nu +1) \right] H_\nu^{(1)}(-p c_s\tau),~~
\nu = \frac{3}{2} + \frac{\eta}{2}+ \epsilon,
\end{equation}
which corresponds to the Bunch-Davies vacuum $\ket{0}$ at early time defined by $ \hat{ a }_{ \mathbf{p} } \ket{0} = 0 $. In general, solution of the MS equation is a linear combination of Hankel function of the first kind $H_\nu^{(1)}(-p c_s\tau)$ and second kind $H_\nu^{(2)}(-p c_s\tau)$. Because we are looking for solution which goes as Bunch-Davies mode function $\exp(-i p c_s \tau)$ at early times $\tau \rightarrow -\infty$, the coefficient of $H_\nu^{(2)}(-p c_s\tau)$ vanishes.

Observable quantities of cosmological perturbations are their correlation function evaluated at the end of inflation. The two-point function of curvature perturbation and power spectrum at a late time during inflation $\tau_0$ are defined as
\begin{gather}
\left\langle \zeta(\mathbf{p}) \zeta(\mathbf{q}) \right\rangle = (2 \pi)^3 \delta^3(\mathbf{p} + \mathbf{q}) \left\langle \! \left\langle \zeta(\mathbf{p}) \zeta(-\mathbf{p}) \right\rangle \! \right\rangle , \\
\Delta^2_s(p) \equiv \frac{p^3}{2 \pi^2} \left\langle \! \left\langle \zeta(\mathbf{p}) \zeta(-\mathbf{p}) \right\rangle \! \right\rangle, 
\end{gather}
and take a limit $\tau_0 \rightarrow 0$, when all the relevant modes are in the superhorizon regime and the observed power spectrum is evaluated. Here
the bracket denotes the VEV,  $\langle \cdots \rangle = \bra{0} \cdots \ket{0}$, and $\Delta^2_s(p)$ is the power spectrum multiplied by the phase space density. 

To obtain the late time limit of curvature perturbation, we need to know the explicit time dependence of $z$. It is given by 
\begin{equation}
\frac{z}{z_H} = \left( \frac{\tau}{\tau_H}  \right)^{- \left( 1 + \epsilon + \frac{\eta}{2} \right) },
\end{equation}
where $\tau_H$ is an arbitrary time reference and $z_H = z(\tau_H)$. Choosing $\tau_H$ as sound horizon crossing time $c_s k = aH$ so
\begin{align}
\tau_H &= - \left[ \frac{1}{aH (1-\epsilon)} \right]_H = - \frac{1}{c_s k(1-\epsilon)_H}, \label{tauh} \\
z_H &= \left( \frac{a}{c_s} \sqrt{2\epsilon} \right)_H = \left( \frac{k}{ H} \sqrt{2\epsilon} \right)_H,
\end{align}
then $z$ as a function of $\tau$ is
\begin{equation}
\label{ztau}
z = \left( \frac{k}{ H} \sqrt{2\epsilon} \right)_H \left[ -c_s k\tau (1-\epsilon)_H \right]^{-\left( 1 + \epsilon + \frac{\eta}{2} \right)}.
\end{equation}
Hence, up to the lowest order in the slow-roll parameters, the mode function is
\begin{equation}
\zeta_p(\tau) = \frac{v_p (\tau)}{zM_{\mathrm{pl}}} = \left( \frac{H^2}{4M_{\mathrm{pl}}^2 \epsilon c_s} \right)_H^{\frac{1}{2}} \frac{e^{-ic_s p \tau}}{p^{3/2}} (1 + i c_s p \tau), \label{modefunction}
\end{equation}
and the power spectrum is
\begin{equation}
\label{spower}
\Delta^2_{s(0)}(p) = \left( \frac{H^2}{8 \pi^2 M_{\mathrm{pl}}^2 c_s \epsilon} \right)_H = \Delta^2_{s(0)}(p_*) \left( \frac{p}{p_*} \right)^{n_s - 1}.
\end{equation}
Here subscripts $(0)$ denotes tree-level contribution and $p_*$ is an arbitrary pivot momentum. The power spectrum is almost scale invariant with the deviation parametrized by the spectral index
\begin{equation}
n_s - 1 = \frac{\md \log \Delta^2_{s(0)}}{\md \log p} = - 2\epsilon - \eta
\end{equation}
Indeed, there is a weak momentum dependence because of time dependence of the quantities in the parenthesis at the horizon crossing of each mode. 

\subsection{Third-order Action}
Expanding more until third-order, in the limit of $c_s \ll 1$, we get \cite{Chen:2009bc,Chen:2013aj,Seery:2005wm}
\begin{equation}
S_{\mathrm{int}} = \int \md t ~\md^3 x \left[ - \frac{2 \lambda}{H^3} a^3 \dot{\zeta}^3 + \frac{\epsilon M_{\mathrm{pl}}^2}{ H c_s^2 } a \dot{\zeta} (\partial_i \zeta)^2 \right], 
\end{equation}
where $\lambda = X^2 P_{,XX} + (2/3) X^3 P_{,XXX}$. Such third-order action can be regarded as cubic self-interaction term of the scalar perturbation. The interaction terms are also predicted by the effective field theory of inflation \cite{Cheung:2007st}. To compare with observational constraint \cite{Akrami:2019izv}, we have to reparametrize $\lambda$ to $\tilde{c}_3$ as
\begin{equation}
\lambda = - \frac{\epsilon M_{\mathrm{pl}}^2 H^2}{3 c_s^4} \tilde{c}_3.
\end{equation}
Therefore, the third-order action becomes
\begin{equation}
S_{\mathrm{int}} = \int \md \tau ~\md^3 x ~\frac{\epsilon \mpl^2}{H c_s^2} \left[ \frac{2 \tilde{c}_3}{3 c_s^2} a^4 \dot{\zeta}^3 + a^2 \dot{\zeta} (\partial_i \zeta)^2 \right],  \label{sint}
\end{equation}
where we define Lagrangian density $\mathcal{L}_{\mathrm{int}}$ as the integrand of \eqref{sint}. Although $\mathcal{L}_{\mathrm{int}}$ contain time derivative of the field, the Hamiltonian density is still given by $\mathcal{H}_{\mathrm{int}} = -\mathcal{L}_{\mathrm{int}}$ as shown by \cite{Adshead:2008gk, Bartolo:2010bu}. The interaction Hamiltonian reads
\begin{equation}
H_{\mathrm{int}}(\tau) = - \int \md^3 x ~\frac{\epsilon \mpl^2}{H c_s^2} \left[ \frac{2 \tilde{c}_3}{3 c_s^2} a^4 \dot{\zeta}^3 + a^2 \dot{\zeta} (\partial_i \zeta)^2 \right].
\end{equation}
Note that the dot still denotes derivative with respect to $t$. We define $H_{\mathrm{int}}$ as function of $\tau$ for use in the following section.

\section{One-Loop Correction}
We are interested in the statistic of fluctuations in the CMB. To calculate higher-order correction, we use the in-in perturbation theory which gives the expectation value of an operator $\mathcal{O}(\tau)$ as 
\begin{equation}
\langle \mathcal{O(\tau)} \rangle =  \left\langle \left[ \bar{\mathrm{T}} \exp \left( i \int_{-\infty}^{\tau} \mathrm{d}\tau' H_{\mathrm{int}}(\tau') \right) \right] \mathcal{\hat{O}} (\tau) \left[ \mathrm{T} \exp \left( -i \int_{-\infty}^{\tau} \mathrm{d\tau'} H_{\mathrm{int}}(\tau') \right) \right] \right\rangle,
\end{equation}
where $\mathrm{T}$ and $\bar{\mathrm{T}}$ denote time and antitime ordering. Because we are interested in the higher-order correction of two-point correlation, the relevant operator is $\zeta(\mathbf{p}) \zeta(-\mathbf{p})$ evaluated at $\tau=\tau_0 ~(\rightarrow 0)$. Its first-order expansion vanishes, yielding an odd-point correlation function. 
Second-order expansion of the perturbation theory reads
\begin{gather}
\langle \mathcal{O(\tau)} \rangle = \langle \mathcal{O(\tau)} \rangle_{(0,2)}^\dagger + \langle \mathcal{O(\tau)} \rangle_{(1,1)} + \langle \mathcal{O(\tau)} \rangle_{(0,2)}, \nonumber\\
\langle \mathcal{O(\tau)} \rangle_{(1,1)} = \int_{-\infty}^{\tau} \mathrm{d}\tau_1 \int_{-\infty}^{\tau} \mathrm{d}\tau_2 \left\langle H_{\mathrm{int}}(\tau_1) \mathcal{\hat{O}} (\tau) H_{\mathrm{int}}(\tau_2) \right\rangle, \\
\langle \mathcal{O(\tau)} \rangle_{(0,2)} = - \int_{-\infty}^{\tau} \mathrm{d}\tau_1 \int_{-\infty}^{\tau_1} \mathrm{d}\tau_2 \left\langle \mathcal{\hat{O}} (\tau) H_{\mathrm{int}}(\tau_1) H_{\mathrm{int}}(\tau_2) \right\rangle.
\end{gather}
They are proportional to the product of two interaction Hamiltonian. Because each interaction Hamiltonian has two interaction terms, namely $\dot{\zeta}^3$ and $\dot{\zeta} (\partial \zeta)^2$, multiplication of them leads to four terms. If we label $\dot{\zeta}^3$ and $\dot{\zeta} (\partial \zeta)^2$ as $a$ and $b$, respectively, the four terms are $[aa]$, $[bb]$, $[ab]$, and $[ba]$.

We start our calculation from the $(1,1)$ contribution of $[aa]$ term that is given by
\begin{align}
\left\langle \zeta(\bp) \zeta(-\bp) \right\rangle_{(1,1)}^{[aa]} = & M_{\mathrm{pl}}^4 \left( \frac{2 \tilde{c}_3}{3 c_s^2} \right)^2 \int_{-\infty}^0 \md \tau_1 \frac{a^4(\tau_1) \epsilon(\tau_1)}{H(\tau_1) c_s^2} \int_{-\infty}^0 \md \tau_2 \frac{a^4(\tau_2) \epsilon(\tau_2)}{H(\tau_2) c_s^2}  \nonumber\\
& \int \prod_{a = 1}^6 \left[ \frac{\md^3 k_a}{(2\pi)^3} \right] \delta^3(\bk_1+\bk_2+\bk_3) ~ \delta^3(\bk_4+\bk_5+\bk_6) \nonumber\\
& \left\langle  \dot{\zeta}(\bk_1,\tau_1) \dot{\zeta}(\bk_2,\tau_1) \dot{\zeta}(\bk_3,\tau_1) \zeta(\bp) \zeta(-\bp) \dot{\zeta}(\bk_4,\tau_2) \dot{\zeta}(\bk_5,\tau_2) \dot{\zeta}(\bk_6,\tau_2) \right\rangle. 
\end{align}
For small loop momentum $k \ll p$, we find
\begin{align}
\left\langle \zeta(\bp) \zeta(-\bp) \right\rangle_{(1,1)}^{[aa]} = \left( \frac{2 \tilde{c}_3}{3 c_s^2} \right)^2 \zeta_p(\tau_0) \zeta_p(\tau_0) & \int_{-\infty}^0 \md \tau_1 ~\tau_1^2 \int_{-\infty}^0 \md \tau_2 ~\tau_2^2 e^{-2i c_s p (\tau_1-\tau_2)} \nonumber\\
& \int \frac{\md^3 k}{(2\pi)^3} \frac{k p^2 c_s^6}{2} \left( \frac{H^2}{4 M_{\mathrm{pl}}^2 \epsilon c_s} \right)_H. 
\end{align}
In the limit of loop momentum $k \rightarrow 0$, the momentum integration becomes
\begin{equation}
\int \frac{\md^3 k}{(2\pi)^3} k \rightarrow 0.
\end{equation}
Thanks to the derivative self-interaction, there is no infrared divergence in the one-loop correction. The other terms, $[bb]$, $[ab]$, and $[ba]$, also generate the same loop momentum integration in the limit of small loop momentum. The $(0,2)$ contribution has a slightly different time integration with the same momentum integration. 

Hence, the loop correction comes from other domain, namely, $k > p$. In this domain, the divergent integral comes from the $(0,2)$ contribution suffers from a divergence, while the $(1,1)$ contribution converges in a similar way to \cite{Senatore:2009cf}. The $(0,2)$ contribution of $[aa]$ term is given by
\begin{align}
\left\langle \zeta(\bp) \zeta(-\bp) \right\rangle_{(0,2)}^{[aa]} = & -M_{\mathrm{pl}}^4 \left( \frac{2 \tilde{c}_3}{3 c_s^2} \right)^2 \int_{-\infty}^0 \md \tau_1 \frac{a^4(\tau_1) \epsilon(\tau_1)}{H(\tau_1) c_s^2} \int_{-\infty}^{\tau_1} \md \tau_2 \frac{a^4(\tau_2) \epsilon(\tau_2)}{H(\tau_2) c_s^2}  \nonumber\\
& \int \prod_{a = 1}^6 \left[ \frac{\md^3 k_a}{(2\pi)^3} \right] \delta^3(\bk_1+\bk_2+\bk_3) ~ \delta^3(\bk_4+\bk_5+\bk_6) \nonumber\\
&  \left\langle \zeta(\bp) \zeta(-\bp) \dot{\zeta}(\bk_1,\tau_1) \dot{\zeta}(\bk_2,\tau_1) \dot{\zeta}(\bk_3,\tau_1) \dot{\zeta}(\bk_4,\tau_2) \dot{\zeta}(\bk_5,\tau_2) \dot{\zeta}(\bk_6,\tau_2) \right\rangle. 
\end{align}
Performing Wick contraction, it becomes
\begin{align}
\left\langle \! \left\langle \zeta(\mathbf{p}) \zeta(-\mathbf{p}) \right\rangle \! \right\rangle_{(0,2)}^{[aa]} = & -M_{\mathrm{pl}}^4 \left( \frac{2 \tilde{c}_3}{3 c_s^2} \right)^2 \zeta_p(\tau_0) \zeta_p(\tau_0) \int_{-\infty}^0 \md \tau_1 \frac{a^4(\tau_1) \epsilon(\tau_1)}{H(\tau_1) c_s^2} \int_{-\infty}^{\tau_1} \md \tau_2 \frac{a^4(\tau_2) \epsilon(\tau_2)}{H(\tau_2) c_s^2} \nonumber\\
& 36 \int \frac{\md^3k}{(2\pi)^3} \dot{\zeta}_k(\tau_1) \dot{\zeta}_k^*(\tau_2) \dot{\zeta}_p^*(\tau_1) \dot{\zeta}_p^*(\tau_2) \dot{\zeta}_q(\tau_1) \dot{\zeta}_q^*(\tau_2), \label{eqaa}
\end{align}
where $\bq = \bk -\bp$. After substituting mode function \eqref{modefunction} and performing time integration, we are left with integration over loop momentum $k$ of a complicated function. The time integration is performed by adding a small $i \epsilon$ prescription to $\tau \rightarrow \tau (1 + i \epsilon)$, so that the integral at early time $\tau \rightarrow -\infty$ converges. Because the integration domain is $k>p$, we can approximate that complicated function by expanding $p \ll k$. As a result, we will get a function with some different power of $k$. For cosmological interest, the most important term is the one with the integrand proportional to $k^{-3}$ up to the slow-roll parameters. Other terms with power of integrand higher than $k^{-3}$ lead to positive power of $k$ after integration. These terms will generate polynomial divergences and should be regularized and renormalized in the same way as QFT in the flat spacetime. Extracting terms with integrand approximately proportional to $ k^{-3}$ yields
\begin{align}
\left\langle \! \left\langle \zeta(\mathbf{p}) \zeta(-\mathbf{p}) \right\rangle \! \right\rangle_{(0,2)}^{[aa]} &\cong \frac{3}{40} \left(\frac{2 \tilde{c}_3}{3 c_s^2} \right)^2 \left\langle \! \left\langle \zeta(\mathbf{p}) \zeta(-\mathbf{p}) \right\rangle \! \right\rangle_{(0)} \int_p^\infty \md k ~k^2 \frac{1}{k^3} \left( \frac{H^2}{8 \pi^2 M_{\mathrm{pl}}^2 \epsilon c_s} \right)_H \nonumber\\
&= \frac{3}{40} \left( \frac{2 \tilde{c}_3}{3 c_s^2} \right)^2  \left\langle \! \left\langle \zeta(\mathbf{p}) \zeta(-\mathbf{p}) \right\rangle \! \right\rangle_{(0)} \int_p^\infty \md k ~k^2 \frac{1}{k^3} \Delta^2_{s(0)}(k_*) \left( \frac{k}{k_*} \right)^{n_s-1}.
\end{align}
The detailed calculation is shown in appendix A. The $(0,2)$ contribution of $[ab]$ term is given by
\begin{align}
\left\langle \zeta(\bp) \zeta(-\bp) \right\rangle_{(0,2)}^{[ab]} = & -M_{\mathrm{pl}}^4 \left( \frac{2 \tilde{c}_3}{3 c_s^2} \right) \int_{-\infty}^0 \md \tau_1 \frac{a^4(\tau_1) \epsilon(\tau_1)}{H(\tau_1) c_s^2} \int_{-\infty}^{\tau_1} \md \tau_2 \frac{a^2(\tau_2) \epsilon(\tau_2)}{H(\tau_2) c_s^2}  \nonumber\\
& \int \prod_{a = 1}^6 \left[ \frac{\md^3 k_a}{(2\pi)^3} \right] \delta^3(\bk_1+\bk_2+\bk_3) ~ \delta^3(\bk_4+\bk_5+\bk_6) (\bk_5 \cdot \bk_6) \nonumber\\
&  \left\langle \zeta(\bp) \zeta(-\bp) \dot{\zeta}(\bk_1,\tau_1) \dot{\zeta}(\bk_2,\tau_1) \dot{\zeta}(\bk_3,\tau_1) \dot{\zeta}(\bk_4,\tau_2) \zeta(\bk_5,\tau_2) \zeta(\bk_6,\tau_2) \right\rangle. 
\end{align}
Performing Wick contraction, it becomes
\begin{align}
\left\langle \! \left\langle \zeta(\mathbf{p}) \zeta(-\mathbf{p}) \right\rangle \! \right\rangle_{(0,2)}^{[ab]} = & -M_{\mathrm{pl}}^4 \left( \frac{2 \tilde{c}_3}{3 c_s^2} \right) \zeta_p(\tau_0) \zeta_p(\tau_0) \int_{-\infty}^0 \md \tau_1 \frac{a^4(\tau_1) \epsilon(\tau_1)}{H(\tau_1) c_s^2} \int_{-\infty}^{\tau_1} \md \tau_2 \frac{a^2(\tau_2) \epsilon(\tau_2)}{H(\tau_2) c_s^2} \nonumber\\
&  \int \frac{\md^3k}{(2\pi)^3} \left[ -12 (\bk \cdot \bq) \dot{\zeta}_p^*(\tau_1) \dot{\zeta}_p^*(\tau_2) \dot{\zeta}_k(\tau_1) \zeta_k^*(\tau_2) \dot{\zeta}_q(\tau_1) \zeta_q^*(\tau_2) \right. \nonumber\\
& \left. \hphantom{\int \frac{\md^3k}{(2\pi)^3}} -8 (\bp \cdot \bk) \dot{\zeta}_q(\tau_1) \dot{\zeta}_q^*(\tau_2) \dot{\zeta}_k(\tau_1) \zeta_k^*(\tau_2) \dot{\zeta}_p^*(\tau_1) \zeta_p^*(\tau_2)  \right. \nonumber\\
& \left. \hphantom{\int \frac{\md^3k}{(2\pi)^3}} + 16 (\bp \cdot \bq) \dot{\zeta}_k(\tau_1) \dot{\zeta}_k^*(\tau_2) \dot{\zeta}_q(\tau_1) \zeta_q^*(\tau_2) \dot{\zeta}_p^*(\tau_1) \zeta_p^*(\tau_2)  \right]. \label{eqab}
\end{align}
The $(0,2)$ contribution of $[ba]$ term is given by
\begin{align}
\left\langle \zeta(\bp) \zeta(-\bp) \right\rangle_{(0,2)}^{[ba]} = & -M_{\mathrm{pl}}^4  \left( \frac{2 \tilde{c}_3}{3 c_s^2} \right) \int_{-\infty}^0 \md \tau_1 \frac{a^2(\tau_1) \epsilon(\tau_1)}{H(\tau_1) c_s^2} \int_{-\infty}^{\tau_1} \md \tau_2 \frac{a^4(\tau_2) \epsilon(\tau_2)}{H(\tau_2) c_s^2}  \nonumber\\
& \int \prod_{a = 1}^6 \left[ \frac{\md^3 k_a}{(2\pi)^3} \right] \delta^3(\bk_1+\bk_2+\bk_3) ~ \delta^3(\bk_4+\bk_5+\bk_6) (\bk_2 \cdot \bk_3) \nonumber\\
&  \left\langle \zeta(\bp) \zeta(-\bp) \dot{\zeta}(\bk_1,\tau_1) \zeta(\bk_2,\tau_1) \zeta(\bk_3,\tau_1) \dot{\zeta}(\bk_4,\tau_2) \dot{\zeta}(\bk_5,\tau_2) \dot{\zeta}(\bk_6,\tau_2) \right\rangle.
\end{align}
Performing Wick contraction, it becomes
\begin{align}
\left\langle \! \left\langle \zeta(\mathbf{p}) \zeta(-\mathbf{p}) \right\rangle \! \right\rangle_{(0,2)}^{[ba]} = & -M_{\mathrm{pl}}^4 \left( \frac{2 \tilde{c}_3}{3 c_s^2} \right) \zeta_p(\tau_0) \zeta_p(\tau_0) \int_{-\infty}^0 \md \tau_1 \frac{a^2(\tau_1) \epsilon(\tau_1)}{H(\tau_1) c_s^2} \int_{-\infty}^{\tau_1} \md \tau_2 \frac{a^4(\tau_2) \epsilon(\tau_2)}{H(\tau_2) c_s^2} \nonumber\\
&  \int \frac{\md^3k}{(2\pi)^3} \left[ 24 (\bp \cdot \bq) \dot{\zeta}_k(\tau_1) \dot{\zeta}_k^*(\tau_2) \zeta_p^*(\tau_1) \dot{\zeta}_p^*(\tau_2) \zeta_q(\tau_1) \dot{\zeta}_q^*(\tau_2) \right. \nonumber\\
& \left. \hphantom{\int \frac{\md^3k}{(2\pi)^3}} +12 (\bk \cdot \bq) \dot{\zeta}_p^*(\tau_1) \dot{\zeta}_p^*(\tau_2) \zeta_k(\tau_1) \dot{\zeta}_k^*(\tau_2) \zeta_q(\tau_1) \dot{\zeta}_q^*(\tau_2)  \right]. \label{eqba}
\end{align}
Extracting terms with integrand approximately proportional to $k^{-3}$, we get the same result for $[ab]$ and $[ba]$. The detailed calculation is shown in appendix B. Summing them yields
\begin{align}
2 \left\langle \! \left\langle \zeta(\mathbf{p}) \zeta(-\mathbf{p}) \right\rangle \! \right\rangle_{(0,2)}^{[ab]} &\cong \frac{9}{40 c_s^2} \left(\frac{2 \tilde{c}_3}{3 c_s^2} \right) \left\langle \! \left\langle \zeta(\mathbf{p}) \zeta(-\mathbf{p}) \right\rangle \! \right\rangle_{(0)} \int_p^\infty \md k ~k^2 \frac{1}{k^3} \left( \frac{H^2}{8 \pi^2 M_{\mathrm{pl}}^2 \epsilon c_s} \right)_H \nonumber\\
&= \frac{9}{40 c_s^2} \left( \frac{2 \tilde{c}_3}{3 c_s^2} \right) \left\langle \! \left\langle \zeta(\mathbf{p}) \zeta(-\mathbf{p}) \right\rangle \! \right\rangle_{(0)} \int_p^\infty \md k ~k^2 \frac{1}{k^3} \Delta^2_{s(0)}(k_*) \left( \frac{k}{k_*} \right)^{n_s-1}. 
\end{align}
The $(0,2)$ contribution of $[bb]$ term is given by
\begin{align}
\left\langle \zeta(\bp) \zeta(-\bp) \right\rangle_{(0,2)}^{[bb]} = & -M_{\mathrm{pl}}^4  \int_{-\infty}^0 \md \tau_1 \frac{a^2(\tau_1) \epsilon(\tau_1)}{H(\tau_1) c_s^2} \int_{-\infty}^{\tau_1} \md \tau_2 \frac{a^2(\tau_2) \epsilon(\tau_2)}{H(\tau_2) c_s^2}  \nonumber\\
& \int \prod_{a = 1}^6 \left[ \frac{\md^3 k_a}{(2\pi)^3} \right] \delta^3(\bk_1+\bk_2+\bk_3) ~ \delta^3(\bk_4+\bk_5+\bk_6) (\bk_2 \cdot \bk_3) (\bk_5 \cdot \bk_6) \nonumber\\
&  \left\langle \zeta(\bp) \zeta(-\bp) \dot{\zeta}(\bk_1,\tau_1) \zeta(\bk_2,\tau_1) \zeta(\bk_3,\tau_1) \dot{\zeta}(\bk_4,\tau_2) \zeta(\bk_5,\tau_2) \zeta(\bk_6,\tau_2) \right\rangle.  \label{start}
\end{align}
Performing Wick contraction, it becomes
\begin{equation}
\left\langle \! \left\langle \zeta(\mathbf{p}) \zeta(-\mathbf{p}) \right\rangle \! \right\rangle_{(0,2)}^{[bb]} = -M_{\mathrm{pl}}^4 \zeta_p(\tau_0) \zeta_p(\tau_0) \int_{-\infty}^0 \md \tau_1 \frac{a^2(\tau_1) \epsilon(\tau_1)}{H(\tau_1) c_s^2} \int_{-\infty}^{\tau_1} \md \tau_2 \frac{a^2(\tau_2) \epsilon(\tau_2)}{H(\tau_2) c_s^2}  f(p, \tau_1, \tau_2), \nonumber\\
\end{equation}
\begin{align}
f(p, \tau_1, \tau_2) = \int \frac{\md^3k}{(2\pi)^3}  & \left[ 8(\bp \cdot \bq)^2  \dot{\zeta}_k(\tau_1) \dot{\zeta}^*_k(\tau_2) \zeta_p^*(\tau_1) \zeta_p^*(\tau_2) \zeta_q(\tau_1) \zeta_q^*(\tau_2)  \right. \nonumber\\
& \left. -8 (\bp \cdot \bk) (\bp \cdot \bq) \dot{\zeta}_k(\tau_1) \dot{\zeta}^*_q(\tau_2) \zeta_p^*(\tau_1) \zeta_p^*(\tau_2) \zeta_q(\tau_1) \zeta_k^*(\tau_2) \right. \nonumber\\
& \left.  -8 (\bp \cdot \bq) (\bk \cdot \bq) \dot{\zeta}_k(\tau_1) \dot{\zeta}^*_p(\tau_2) \zeta_p^*(\tau_1) \zeta_k^*(\tau_2) \zeta_q(\tau_1) \zeta_q^*(\tau_2)  \right. \nonumber\\
& \left. -8 (\bp \cdot \bq) (\bk \cdot \bq) \dot{\zeta}^*_p(\tau_1) \dot{\zeta}^*_k(\tau_2) \zeta_k(\tau_1) \zeta_p^*(\tau_2) \zeta_q(\tau_1) \zeta_q^*(\tau_2) \right. \nonumber\\
& \left.  + 4(\bk \cdot \bq)^2  \dot{\zeta}^*_p(\tau_1) \dot{\zeta}^*_p(\tau_2) \zeta_k(\tau_1) \zeta_k^*(\tau_2) \zeta_q(\tau_1) \zeta_q^*(\tau_2) \right]. \label{eqbb} 
\end{align}
Extracting terms with integrand approximately proportional to $k^{-3}$ yields
\begin{align}
\left\langle \! \left\langle \zeta(\mathbf{p}) \zeta(-\mathbf{p}) \right\rangle \! \right\rangle_{(0,2)}^{[bb]} &\cong \frac{51}{80 c_s^4}  \left\langle \! \left\langle \zeta(\mathbf{p}) \zeta(-\mathbf{p}) \right\rangle \! \right\rangle_{(0)} \int_p^\infty \md k ~k^2 \frac{1}{k^3} \left( \frac{H^2}{8 \pi^2 M_{\mathrm{pl}}^2 \epsilon c_s} \right)_H \nonumber\\
&= \frac{51}{80 c_s^4}  \left\langle \! \left\langle \zeta(\mathbf{p}) \zeta(-\mathbf{p}) \right\rangle \! \right\rangle_{(0)} \int_p^\infty \md k ~k^2 \frac{1}{k^3} \Delta^2_{s(0)}(k_*) \left( \frac{k}{k_*} \right)^{n_s-1}. \label{loop}
\end{align}
The detailed calculation is shown in appendix C. Summing all terms $[aa]$, $[bb]$, $[ab]$, and $[ba]$, the total $(0,2)$ contribution to the one-loop correction is
\begin{align}
\left\langle \! \left\langle \zeta(\mathbf{p}) \zeta(-\mathbf{p}) \right\rangle \! \right\rangle_{(0,2)} = \left\langle \! \left\langle \zeta(\mathbf{p}) \zeta(-\mathbf{p}) \right\rangle \! \right\rangle_{(0)} & \left[ \frac{51}{80 c_s^4} + \frac{3}{20 c_s^2} \left( \frac{ \tilde{c}_3 }{ c_s^2 } \right) + \frac{1}{30} \left( \frac{ \tilde{c}_3 }{ c_s^2 } \right)^2 \right]  \nonumber\\
& \times \int_p^\infty \md k ~k^2 \frac{1}{k^3} \Delta^2_{s(0)}(k_*) \left( \frac{k}{k_*} \right)^{n_s-1}. 
\end{align}
For $n_s < 1$, which is the case in our Universe, the momentum integration converges as \cite{Kristiano:2021urj}
\begin{equation}
\left\langle \! \left\langle \zeta(\mathbf{p}) \zeta(-\mathbf{p}) \right\rangle \! \right\rangle_{(0,2)} = \left\langle \! \left\langle \zeta(\mathbf{p}) \zeta(-\mathbf{p}) \right\rangle \! \right\rangle_{(0)} \frac{\Delta^2_{s(0)}(p) }{1 - n_s} \left[ \frac{51}{80 c_s^4} + \frac{3}{20 c_s^2} \left( \frac{ \tilde{c}_3 }{ c_s^2 } \right) + \frac{1}{30} \left( \frac{ \tilde{c}_3 }{ c_s^2 } \right)^2 \right]. 
\end{equation}
Hence, adding the $(2,0)$ contribution, the total one-loop correction to the power spectrum is
\begin{equation}
\Delta^2_{s(1)}(p) = \frac{[\Delta^2_{s(0)}(p)]^2}{1 - n_s} \left[ \frac{51}{40 c_s^4} + \frac{3}{10 c_s^2} \left( \frac{ \tilde{c}_3 }{ c_s^2 } \right) + \frac{1}{15} \left( \frac{ \tilde{c}_3 }{ c_s^2 } \right)^2 \right]. 
\end{equation}
Perturbativity requirement $\Delta^2_{s(1)}(p) \ll \Delta^2_{s(0)}(p)$ reads
\begin{equation}
\frac{\Delta^2_{s(0)}(p)}{1 - n_s} \left[ \frac{51}{40 c_s^4} + \frac{3}{10 c_s^2} \left( \frac{ \tilde{c}_3 }{ c_s^2 } \right) + \frac{1}{15} \left( \frac{ \tilde{c}_3 }{ c_s^2 } \right)^2 \right] \ll 1. \label{perturbative1}
\end{equation}
The total power spectrum as a summation of tree-level and one-loop correction is 
\begin{equation}
\Delta^2_{s}(p) = \Delta^2_{s(0)}(p) \left\lbrace 1 + \frac{\Delta^2_{s(0)}(p)}{1 - n_s} \left[ \frac{51}{40 c_s^4} + \frac{3}{10 c_s^2} \left( \frac{ \tilde{c}_3 }{ c_s^2 } \right) + \frac{1}{15} \left( \frac{ \tilde{c}_3 }{ c_s^2 } \right)^2 \right] \right\rbrace. 
\label{total}
\end{equation}
We can write the explicit momentum dependence of total power spectrum as 
\begin{equation}
\Delta^2_{s}(p) = \Delta^2_{s}(p_*) \left( \frac{p}{p_*} \right)^{N_s - 1},
\end{equation}
where
\begin{align}
N_s - 1 & = \frac{\md \log \Delta^2_s}{\md \log p} \nonumber\\
& =  \frac{\md \log \Delta^2_{s(0)}}{\md \log p} +  \frac{\md}{\md \log p} \log \left\lbrace 1 + \frac{\Delta^2_{s(0)}}{1 - n_s} \left[ \frac{51}{40 c_s^4} + \frac{3}{10 c_s^2} \left( \frac{ \tilde{c}_3 }{ c_s^2 } \right) + \frac{1}{15} \left( \frac{ \tilde{c}_3 }{ c_s^2 } \right)^2 \right] \right\rbrace \nonumber\\
& \approx \frac{\md \log \Delta^2_{s(0)}}{\md \log p} + \frac{\md}{\md \log p} \left\lbrace \frac{\Delta^2_{s(0)}}{1 - n_s} \left[ \frac{51}{40 c_s^4} + \frac{3}{10 c_s^2} \left( \frac{ \tilde{c}_3 }{ c_s^2 } \right) + \frac{1}{15} \left( \frac{ \tilde{c}_3 }{ c_s^2 } \right)^2 \right] \right\rbrace \nonumber\\
& = n_s - 1 - \Delta^2_{s(0)} \left[ \frac{51}{40 c_s^4} + \frac{3}{10 c_s^2} \left( \frac{ \tilde{c}_3 }{ c_s^2 } \right) + \frac{1}{15} \left( \frac{ \tilde{c}_3 }{ c_s^2 } \right)^2 \right]. \label{Ns}
\end{align}

One might wonder whether we can have a smooth exact scale-invariant limit $n_s = 1$ of the total power spectrum \eqref{total}. For the exactly scale-invariant case, one-loop correction is proportional to
\begin{equation}
\int \frac{\md^3 k}{(2 \pi)^3 k^3} 
\end{equation}
Performing dynamical dimensional regularization \cite{Green:2020txs,Cohen:2020php,Burgess:2009bs}
\begin{align}
\int \frac{\md^3 k}{(2 \pi)^3 k^3} \rightarrow \mu^\epsilon \int \frac{\md^3 k}{(2 \pi)^3 k^{3+\epsilon}} & = \frac{\mu^\epsilon}{2 \pi^2} \int_p^\infty \frac{\md k}{k^{1+\epsilon}} \nonumber\\
& = \frac{1}{2\pi^2 \epsilon} \left( \frac{p}{\mu} \right)^{-\epsilon} \approx \frac{1}{2\pi^2} \left( \frac{1}{\epsilon} - \log \frac{p}{\mu} \right),
\end{align}
where we introduce a very small positive regulator $\epsilon$ and momentum scale $\mu$ to make the integral dimensionless. The one-loop corrected power spectrum is
\begin{equation}
\Delta^2_{s}(p) = \Delta^2_{s(0)} \left\lbrace 1 + \left( \frac{1}{\epsilon} - \log \frac{p}{\mu} \right) \Delta^2_{s(0)} \left[ \frac{51}{40 c_s^4} + \frac{3}{10 c_s^2} \left( \frac{ \tilde{c}_3 }{ c_s^2 } \right) + \frac{1}{15} \left( \frac{ \tilde{c}_3 }{ c_s^2 } \right)^2 \right] \right\rbrace, 
\end{equation}
which contains divergence $1/\epsilon$. To absorb the divergence, we introduce a renormalization factor $\tilde{\Delta}^2_{s}(p) = Z \Delta^2_{s}(p)$ with
\begin{equation}
Z = 1 - \left( \frac{1}{\epsilon} - \log \frac{\tilde{p}}{\mu} \right) \Delta^2_{s(0)} \left[ \frac{51}{40 c_s^4} + \frac{3}{10 c_s^2} \left( \frac{ \tilde{c}_3 }{ c_s^2 } \right) + \frac{1}{15} \left( \frac{ \tilde{c}_3 }{ c_s^2 } \right)^2 \right],
\end{equation}
and perform renormalization:
\begin{equation}
\tilde{\Delta}^2_{s}(p) = \Delta^2_{s(0)} \left\lbrace 1 - \Delta^2_{s(0)} \left[ \frac{51}{40 c_s^4} + \frac{3}{10 c_s^2} \left( \frac{ \tilde{c}_3 }{ c_s^2 } \right) + \frac{1}{15} \left( \frac{ \tilde{c}_3 }{ c_s^2 } \right)^2 \right] \log \frac{p}{\tilde{p}} \right\rbrace. 
\end{equation}
This renormalization fixes the total power spectrum at scale $\tilde{p}$ as $\tilde{\Delta}^2_{s}(\tilde{p}) = \Delta^2_{s(0)}$. Observable $\tilde{\Delta}^2_{s}(p)$ must not depend on arbitrary parameter $\tilde{p}$. It can be realised if $\Delta^2_{s(0)}$ runs as a function of $\tilde{p}$, which satisfies the following differential equation
\begin{align}
0 & = \frac{\partial}{\partial \log \tilde{p}} \log \tilde{\Delta}^2_{s}(p) \nonumber\\
& \approx \frac{\partial}{\partial \log \tilde{p}} \left\lbrace \log \Delta^2_{s(0)} - \Delta^2_{s(0)} \left[ \frac{51}{40 c_s^4} + \frac{3}{10 c_s^2} \left( \frac{ \tilde{c}_3 }{ c_s^2 } \right) + \frac{1}{15} \left( \frac{ \tilde{c}_3 }{ c_s^2 } \right)^2 \right] \log \frac{p}{\tilde{p}} \right\rbrace \nonumber\\
& \approx \frac{\partial \log \Delta^2_{s(0)}}{\partial \log \tilde{p}} + \Delta^2_{s(0)} \left[ \frac{51}{40 c_s^4} + \frac{3}{10 c_s^2} \left( \frac{ \tilde{c}_3 }{ c_s^2 } \right) + \frac{1}{15} \left( \frac{ \tilde{c}_3 }{ c_s^2 } \right)^2 \right], \nonumber\\
\frac{\partial \log \Delta^2_{s(0)}}{\partial \log \tilde{p}} & = - \Delta^2_{s(0)} \left[ \frac{51}{40 c_s^4} + \frac{3}{10 c_s^2} \left( \frac{ \tilde{c}_3 }{ c_s^2 } \right) + \frac{1}{15} \left( \frac{ \tilde{c}_3 }{ c_s^2 } \right)^2 \right].
\end{align}
By identifying $\Delta^2_{s(0)} = \tilde{\Delta}^2_{s}(\tilde{p})$ and redefine $\tilde{p} \rightarrow p$, we get
\begin{equation}
\tilde{\Delta}^2_{s}(p) = \tilde{\Delta}^2_{s}(p_*) \left( \frac{p}{p_*} \right)^{N_s-1},
\end{equation}
where
\begin{equation}
N_s - 1 = - \Delta^2_{s(0)} \left[ \frac{51}{40 c_s^4} + \frac{3}{10 c_s^2} \left( \frac{ \tilde{c}_3 }{ c_s^2 } \right) + \frac{1}{15} \left( \frac{ \tilde{c}_3 }{ c_s^2 } \right)^2 \right]. 
\end{equation}
Therefore, there is a smooth limit in which we recover the momentum dependence of renormalized power spectrum of the exact scale-invariant result, by directly substituting $n_s = 1$ to \eqref{Ns}.

\section{Constraint on Parameter Space}

Perturbativity requirement \eqref{perturbative1} leads to a constraint on parameter space $\tilde{c}_3$ and $c_s$. In Figure 1, we compare our theoretical constraint to observational constraint \cite{Akrami:2019izv}. We know that $\Delta_{s(0)}^2\equiv A_2=2.1\times 10^{-9}$ at pivot scale $k_\ast=0.05~\mathrm{Mpc}^{-1}$ and $n_s=0.9649 \pm 0.0042$ based on $TT$, $TE$, $EE$, low $E$, and lens of Planck 2018 \cite{Akrami:2018odb}. The curve $\Delta^2_{s(1)} = \Delta^2_{s(0)}$ is shown by the red curve. Outside the red curve, perturbativity no longer works and we must not trust perturbation theory anymore. The parameter space that is allowed by Planck 2018 observation is enclosed by the blue curve. We can see that the allowed parameter space is inside the red curve, so the observationally allowed parameters have one-loop correction less than tree-level contribution and perturbativity treatment works.

However, it is important to require strong inequality on the one-loop correction compared to the tree-level contribution, namely $\Delta^2_{s(1)} \ll \Delta^2_{s(0)}$. Hence, we also draw the curve $\Delta^2_{s(1)} = 0.1 \Delta^2_{s(0)}$ and $\Delta^2_{s(1)} = 0.01 \Delta^2_{s(0)}$, shown by the magenta and green curve, respectively. For the region inside the green curve, we can safely neglect one-loop correction because it is very small. For the region enclosed by the magenta and the green curve, we might consider to add one-loop correction, depending on the precision that we want to achieve. For observationally allowed parameter between red and magenta curve, one-loop correction is large enough, greater than $0.1$ times tree-level contribution. Hence, if future observations squeeze the parameter space into the region between red and magenta curve, then considering one-loop correction in the theory will be important to achieve precise cosmology.

\begin{figure}[tbp]
\centering 
\includegraphics[width=\textwidth]{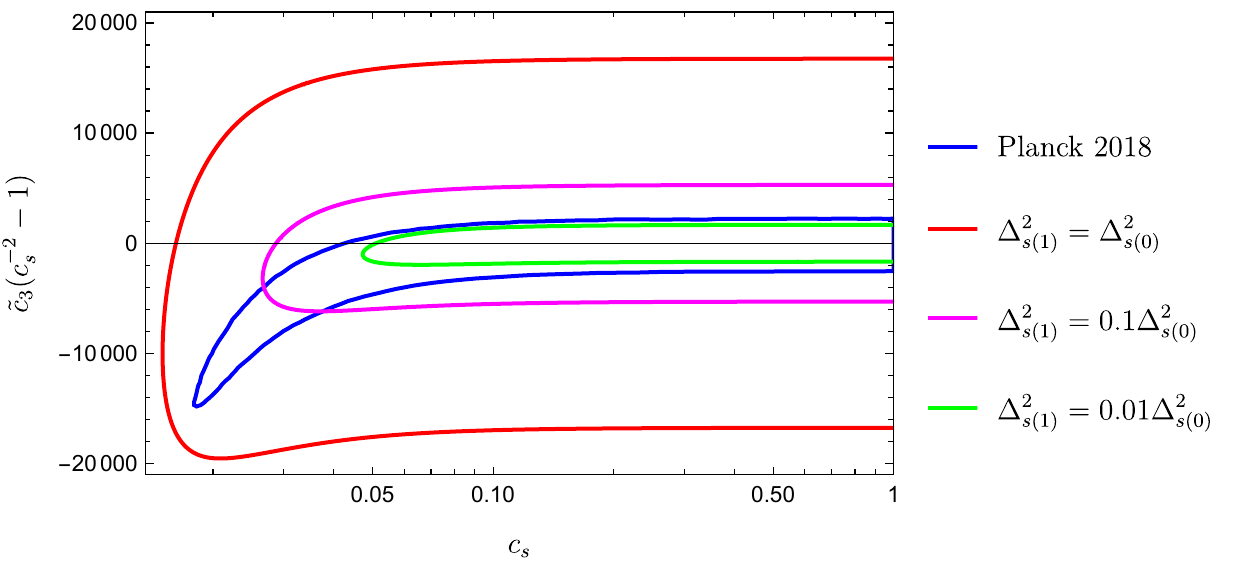}
\caption{\label{fig:i} Comparison between theoretical and observational constraint on $\tilde{c}_3$ and $c_s$ parameter space. Region inside blue curve is Planck 2018 observationally allowed parameters with $99.7\%$ confidence level \cite{Akrami:2019izv}. Theoretical constraint when one-loop correction $\Delta^2_{s(1)}$ equals to tree-level contribution $\Delta^2_{s(0)}$ is shown by the red curve. Another constraint when $\Delta^2_{s(1)} = 0.1\Delta^2_{s(0)}$ and $\Delta^2_{s(1)} = 0.01\Delta^2_{s(0)}$ are shown by the magenta and green curve, respectively.}
\end{figure}

\section{Conclusion}

Perturbativity may not be taken for granted. We showed that it imposes non-trivial constraints to the theory. Requiring one-loop correction to the two-point functions of curvature perturbation to be much smaller than its tree-level contribution leads to an upper bound on primordial non-Gaussianity. We compared this theoretical constraint to observational constraint on non-Gaussian parameter space. Although all observationally allowed parameters have one-loop correction less than tree-level contribution, a quite large region in the parameter space suffers from one-loop correction greater than $0.1$ times tree-level contribution. Therefore, if future observations find non-Gaussianity inside this region, it will be important to consider one-loop correction in the cosmological perturbation theory.

\appendix
\section{Calculation of $\dot{\zeta}^3 \times \dot{\zeta}^3$ term}
In this appendix, we calculate the one-loop correction from $\dot{\zeta}^3 \times \dot{\zeta}^3$ term, that we label as $[aa]$ term. From \eqref{eqaa}, it is given by
\begin{align}
\left\langle \! \left\langle \zeta(\mathbf{p}) \zeta(-\mathbf{p}) \right\rangle \! \right\rangle_{(0,2)}^{[aa]} = & -M_{\mathrm{pl}}^4 \left( \frac{2 \tilde{c}_3}{3 c_s^2} \right)^2 \left\langle \! \left\langle \zeta(\mathbf{p}) \zeta(-\mathbf{p}) \right\rangle \! \right\rangle_{(0)} \int_{-\infty}^0 \md \tau_1 \frac{a^4(\tau_1) \epsilon(\tau_1)}{H(\tau_1) c_s^2} \int_{-\infty}^{\tau_1} \md \tau_2 \frac{a^4(\tau_2) \epsilon(\tau_2)}{H(\tau_2) c_s^2} \nonumber\\
& 36 \int \frac{\md^3k}{(2\pi)^3} \dot{\zeta}_k(\tau_1) \dot{\zeta}_k^*(\tau_2) \dot{\zeta}_p^*(\tau_1) \dot{\zeta}_p^*(\tau_2) \dot{\zeta}_q(\tau_1) \dot{\zeta}_q^*(\tau_2).
\end{align}
Substituting mode function \eqref{modefunction} leads to
\begin{align}
\left\langle \! \left\langle \zeta(\mathbf{p}) \zeta(-\mathbf{p}) \right\rangle \! \right\rangle_{(0,2)}^{[aa]} = & -\frac{9 c_s^6}{4} \left( \frac{2 \tilde{c}_3}{3 c_s^2} \right)^2   \left\langle \! \left\langle \zeta(\mathbf{p}) \zeta(-\mathbf{p}) \right\rangle \! \right\rangle_{(0)} \int \frac{\md^3k}{(2\pi)^3} \left( \frac{H^2}{4 M_{\mathrm{pl}}^2 \epsilon c_s} \right)_H k p q  \nonumber\\
& \int_{-\infty}^0 \md \tau_1 ~\tau_1^2 \int_{-\infty}^{\tau_1} \md \tau_2 ~\tau_2^2 e^{i c_s (p-k-q) \tau_1} e^{i c_s (p+k+q) \tau_2}.  
\end{align}
Then, performing time integration yields
\begin{align}
\left\langle \! \left\langle \zeta(\mathbf{p}) \zeta(-\mathbf{p}) \right\rangle \! \right\rangle_{(0,2)}^{[aa]} = &~ \frac{9 c_s^6}{4} \left( \frac{2 \tilde{c}_3}{3 c_s^2} \right)^2   \left\langle \! \left\langle \zeta(\mathbf{p}) \zeta(-\mathbf{p}) \right\rangle \! \right\rangle_{(0)} \int \frac{\md^3k}{(2\pi)^3} \left( \frac{H^2}{4 M_{\mathrm{pl}}^2 \epsilon c_s} \right)_H k p q  \nonumber\\
& \times \frac{3 k^2+9 k p+6 k q+8 p^2+9 p q+3 q^2}{4c_s^6 p^5 (k+p+q)^3}.
\end{align}
Substituting $q^2 = k^2 - 2 \bp \cdot \bk + p^2$, expanding with respect to the power of $p/k \ll 1$, and extracting terms that have integrand approximately proportional to $k^{-3}$ leads to
\begin{align}
\left\langle \! \left\langle \zeta(\mathbf{p}) \zeta(-\mathbf{p}) \right\rangle \! \right\rangle_{(0,2)}^{[aa]} = &- \frac{9 c_s^6}{4} \left( \frac{2 \tilde{c}_3}{3 c_s^2} \right)^2  \left\langle \! \left\langle \zeta(\mathbf{p}) \zeta(-\mathbf{p}) \right\rangle \! \right\rangle_{(0)}  \nonumber\\
& \int \frac{\md^3k}{(2\pi)^3} \left( \frac{H^2}{4 M_{\mathrm{pl}}^2 \epsilon c_s} \right)_H \frac{1}{c_s^6 p^4 k^3} \left[ \frac{21 s^4}{64} - \frac{11 p^2 s^2}{32} + \frac{p^4}{64} \right],
\end{align}
where $s = - \bp \cdot \hat{\bk}$. Next, substituting $\hat{\bp} \cdot \hat{\bk} = \cos\theta$ and decomposing volume element yields
\begin{align}
\left\langle \! \left\langle \zeta(\mathbf{p}) \zeta(-\mathbf{p}) \right\rangle \! \right\rangle_{(0,2)}^{[aa]} = &- \frac{9}{8} \left( \frac{2 \tilde{c}_3}{3 c_s^2} \right)^2  \left\langle \! \left\langle \zeta(\mathbf{p}) \zeta(-\mathbf{p}) \right\rangle \! \right\rangle_{(0)} \int_p^\infty \md k ~k^2 \frac{1}{k^3} \left( \frac{H^2}{8 \pi^2 M_{\mathrm{pl}}^2 \epsilon c_s} \right)_H  \nonumber\\
& \int_{-1}^1 \md \cos\theta \left[ \frac{21}{64} \cos^4\theta - \frac{11}{32} \cos^2\theta + \frac{1}{64} \right].
\end{align}
We will use the same definition of $s$ and $\theta$ in the next appendix. Finally, after performing angular integration we get
\begin{equation}
\left\langle \! \left\langle \zeta(\mathbf{p}) \zeta(-\mathbf{p}) \right\rangle \! \right\rangle_{(0,2)}^{[aa]} = \frac{3}{40} \left( \frac{2 \tilde{c}_3}{3 c_s^2} \right)^2  \left\langle \! \left\langle \zeta(\mathbf{p}) \zeta(-\mathbf{p}) \right\rangle \! \right\rangle_{(0)} \int_p^\infty \md k ~k^2 \frac{1}{k^3} \left( \frac{H^2}{8 \pi^2 M_{\mathrm{pl}}^2 \epsilon c_s} \right)_H.
\end{equation}

\section{Calculation of $\dot{\zeta}^3 \times \dot{\zeta}(\partial \zeta)^2$ term}
In this appendix, we calculate the one-loop correction from the $\dot{\zeta}^3 \times \dot{\zeta}(\partial \zeta)^2$ cross term, that we label as $[ab]$ and $[ba]$. From \eqref{eqab}, the $[ab]$ term is given by
\begin{align}
\left\langle \! \left\langle \zeta(\mathbf{p}) \zeta(-\mathbf{p}) \right\rangle \! \right\rangle_{(0,2)}^{[ab]} = & -M_{\mathrm{pl}}^4 \left( \frac{2 \tilde{c}_3}{3 c_s^2} \right)  \left\langle \! \left\langle \zeta(\mathbf{p}) \zeta(-\mathbf{p}) \right\rangle \! \right\rangle_{(0)}   \int_{-\infty}^0 \md \tau_1 \frac{a^4(\tau_1) \epsilon(\tau_1)}{H(\tau_1) c_s^2} \int_{-\infty}^{\tau_1} \md \tau_2 \frac{a^2(\tau_2) \epsilon(\tau_2)}{H(\tau_2) c_s^2} \nonumber\\
&  \int \frac{\md^3k}{(2\pi)^3} \left[ -12 (\bk \cdot \bq) \dot{\zeta}_p^*(\tau_1) \dot{\zeta}_p^*(\tau_2) \dot{\zeta}_k(\tau_1) \zeta_k^*(\tau_2) \dot{\zeta}_q(\tau_1) \zeta_q^*(\tau_2) \right. \nonumber\\
& \left. \hphantom{\int \frac{\md^3k}{(2\pi)^3}} -8 (\bp \cdot \bk) \dot{\zeta}_q(\tau_1) \dot{\zeta}_q^*(\tau_2) \dot{\zeta}_k(\tau_1) \zeta_k^*(\tau_2) \dot{\zeta}_p^*(\tau_1) \zeta_p^*(\tau_2)  \right. \nonumber\\
& \left. \hphantom{\int \frac{\md^3k}{(2\pi)^3}} + 16 (\bp \cdot \bq) \dot{\zeta}_k(\tau_1) \dot{\zeta}_k^*(\tau_2) \dot{\zeta}_q(\tau_1) \zeta_q^*(\tau_2) \dot{\zeta}_p^*(\tau_1) \zeta_p^*(\tau_2)  \right].
\end{align}
Substituting mode function \eqref{modefunction} leads to
\begin{align}
\left\langle \! \left\langle \zeta(\mathbf{p}) \zeta(-\mathbf{p}) \right\rangle \! \right\rangle_{(0,2)}^{[ab]} = & - \frac{1}{16}\left( \frac{2 \tilde{c}_3}{3 c_s^2} \right) \left\langle \! \left\langle \zeta(\mathbf{p}) \zeta(-\mathbf{p}) \right\rangle \! \right\rangle_{(0)}   \nonumber\\
& \int_{-\infty}^0 \md \tau_1 ~\tau_1^2 \int_{-\infty}^{\tau_1} \md \tau_2 \int \frac{\md^3k}{(2\pi)^3}  \left( \frac{H^2}{4 M_{\mathrm{pl}}^2 \epsilon c_s} \right)_H  \frac{c_s^2}{p k q}  e^{i c_s (p-k-q) \tau_1} e^{i c_s (p+k+q) \tau_2}  \nonumber\\
& \hphantom{ \int_{-\infty}^0 \md \tau_1 ~\tau_1^2 \int_{-\infty}^{\tau_1} \md \tau_2 } \left[ -12 p^2 (\bk \cdot \bq) (1 - i c_s k \tau_2) (1 - i c_s q \tau_2)   \right. \nonumber\\
& \hphantom{ \int_{-\infty}^0 \md \tau_1 ~\tau_1^2 \int_{-\infty}^{\tau_1} \md \tau_2 } \left. -8q^2 (\bp \cdot \bk) (1 - i c_s p \tau_2) (1- i c_s k \tau_2) \right. \nonumber\\
& \hphantom{ \int_{-\infty}^0 \md \tau_1 ~\tau_1^2 \int_{-\infty}^{\tau_1} \md \tau_2 } \left. + 16 k^2 (\bp \cdot \bq)  (1 - i c_s p \tau_2) (1- i c_s q \tau_2) \right].
\end{align}
Then, performing time integration yields
\begin{align}
\left\langle \! \left\langle \zeta(\mathbf{p}) \zeta(-\mathbf{p}) \right\rangle \! \right\rangle_{(0,2)}^{[ab]} = & - \frac{1}{16 c_s^2}\left( \frac{2 \tilde{c}_3}{3 c_s^2} \right) \left\langle \! \left\langle \zeta(\mathbf{p}) \zeta(-\mathbf{p}) \right\rangle \! \right\rangle_{(0)} \int \frac{\md^3k}{(2\pi)^3}  \left( \frac{H^2}{4 M_{\mathrm{pl}}^2 \epsilon c_s} \right)_H  \frac{1}{p k q}  \nonumber\\
& \frac{(-k)}{2 p^4 (k+p+q)^3} \left[s \left(q^2 \left(98 k^2 p+54 k^3+15 k p^2+16 p^3\right) \right. \right. \nonumber\\
& \left. +3 q \left(71 k^2 p^2+54 k^3 p+12 k^4+36 k p^3+9 p^4\right)  -2 q^4 (9 k+5 p) \right. \nonumber\\
& \left. +p (k+p) \left(37 k^2 p+20 k^3+21 k p^2+6 p^3\right) -3 p q^3 (18 k+5 p)\right) \nonumber\\
& \left. +k p \left(q^2 \left(36 k^2+153 k p+154 p^2\right)  +3 q \left(39 k^2 p+6 k^3+84 k p^2+53 p^3\right) \right. \right. \nonumber\\
& \left. \left. +p (k+p) \left(9 k^2+41 k p+34 p^2\right) +9 q^3 (2 k+5 p) \right) \right].
\end{align}
Substituting $q^2 = k^2 - 2 \bp \cdot \bk + p^2$, expanding with respect to the power of $p/k \ll 1$, and extracting terms that have integrand approximately proportional to $k^{-3}$ leads to
\begin{align}
\left\langle \! \left\langle \zeta(\mathbf{p}) \zeta(-\mathbf{p}) \right\rangle \! \right\rangle_{(0,2)}^{[ab]} &= - \frac{1}{16 c_s^2}\left( \frac{2 \tilde{c}_3}{3 c_s^2} \right) \left\langle \! \left\langle \zeta(\mathbf{p}) \zeta(-\mathbf{p}) \right\rangle \! \right\rangle_{(0)}  \nonumber\\
& \hphantom{=} \int \frac{\md^3k}{(2\pi)^3}  \left( \frac{H^2}{4 M_{\mathrm{pl}}^2 \epsilon c_s} \right)_H \left[ \frac{341 s^4}{16 k^3 p^4} - \frac{357 s^2}{16 k^3 p^2} + \frac{11}{8 k^3} \right] \nonumber\\
& = - \frac{1}{32 c_s^2}\left( \frac{2 \tilde{c}_3}{3 c_s^2} \right) \left\langle \! \left\langle \zeta(\mathbf{p}) \zeta(-\mathbf{p}) \right\rangle \! \right\rangle_{(0)} \int_p^\infty \md k ~k^2 \frac{1}{k^3} \left( \frac{H^2}{8 \pi^2 M_{\mathrm{pl}}^2 \epsilon c_s} \right)_H  \nonumber\\
& \hphantom{=} \int_{-1}^1 \md \cos\theta \left[ \frac{341}{16} \cos^4\theta - \frac{357}{16} \cos^2\theta + \frac{11}{8} \right].
\end{align}
Finally, after performing angular integration we get
\begin{equation}
\left\langle \! \left\langle \zeta(\mathbf{p}) \zeta(-\mathbf{p}) \right\rangle \! \right\rangle_{(0,2)}^{[ab]} = \frac{9}{80 c_s^2}\left( \frac{2 \tilde{c}_3}{3 c_s^2} \right) \left\langle \! \left\langle \zeta(\mathbf{p}) \zeta(-\mathbf{p}) \right\rangle \! \right\rangle_{(0)}  \int_p^\infty \md k ~k^2 \frac{1}{k^3} \left( \frac{H^2}{8 \pi^2 M_{\mathrm{pl}}^2 \epsilon c_s} \right)_H .
\end{equation}
Next, we calculate the $[ba]$ term. From \eqref{eqba}, it is given by
\begin{align}
\left\langle \! \left\langle \zeta(\mathbf{p}) \zeta(-\mathbf{p}) \right\rangle \! \right\rangle_{(0,2)}^{[ba]} = & -M_{\mathrm{pl}}^4 \left( \frac{2 \tilde{c}_3}{3 c_s^2} \right)  \left\langle \! \left\langle \zeta(\mathbf{p}) \zeta(-\mathbf{p}) \right\rangle \! \right\rangle_{(0)}   \int_{-\infty}^0 \md \tau_1 \frac{a^2(\tau_1) \epsilon(\tau_1)}{H(\tau_1) c_s^2} \int_{-\infty}^{\tau_1} \md \tau_2 \frac{a^4(\tau_2) \epsilon(\tau_2)}{H(\tau_2) c_s^2} \nonumber\\
&  \int \frac{\md^3k}{(2\pi)^3} \left[ 24 (\bp \cdot \bq) \dot{\zeta}_k(\tau_1) \dot{\zeta}_k^*(\tau_2) \zeta_p^*(\tau_1) \dot{\zeta}_p^*(\tau_2) \zeta_q(\tau_1) \dot{\zeta}_q^*(\tau_2) \right. \nonumber\\
& \left. \hphantom{\int \frac{\md^3k}{(2\pi)^3}} +12 (\bk \cdot \bq) \dot{\zeta}_p^*(\tau_1) \dot{\zeta}_p^*(\tau_2) \zeta_k(\tau_1) \dot{\zeta}_k^*(\tau_2) \zeta_q(\tau_1) \dot{\zeta}_q^*(\tau_2)  \right].
\end{align}
Substituting mode function \eqref{modefunction} leads to
\begin{align}
\left\langle \! \left\langle \zeta(\mathbf{p}) \zeta(-\mathbf{p}) \right\rangle \! \right\rangle_{(0,2)}^{[ba]} = & - \frac{1}{16} \left( \frac{2 \tilde{c}_3}{3 c_s^2} \right) \left\langle \! \left\langle \zeta(\mathbf{p}) \zeta(-\mathbf{p}) \right\rangle \! \right\rangle_{(0)}   \nonumber\\
& \int_{-\infty}^0 \md \tau_1 \int_{-\infty}^{\tau_1} \md \tau_2 ~\tau_2^2 \int \frac{\md^3k}{(2\pi)^3}  \left( \frac{H^2}{4 M_{\mathrm{pl}}^2 \epsilon c_s} \right)_H  \frac{c_s^2}{p k q}  e^{i c_s (p-k-q) \tau_1} e^{i c_s (p+k+q) \tau_2}  \nonumber\\
& \hphantom{ \int_{-\infty}^0 \md \tau_1 \int_{-\infty}^{\tau_1} \md \tau_2 ~\tau_2^2 } \left[ 24 k^2 (\bp \cdot \bq) (1 - i c_s p \tau_1) (1 + i c_s q \tau_1)   \right. \nonumber\\
& \hphantom{ \int_{-\infty}^0 \md \tau_1 \int_{-\infty}^{\tau_1} \md \tau_2 ~\tau_2^2 } \left. +12 p^2 (\bk \cdot \bq) (1 + i c_s k \tau_1) (1 + i c_s q \tau_1) \right].
\end{align}
Then, performing time integration yields
\begin{align}
\left\langle \! \left\langle \zeta(\mathbf{p}) \zeta(-\mathbf{p}) \right\rangle \! \right\rangle_{(0,2)}^{[ba]} = & - \frac{1}{16 c_s^2}\left( \frac{2 \tilde{c}_3}{3 c_s^2} \right) \left\langle \! \left\langle \zeta(\mathbf{p}) \zeta(-\mathbf{p}) \right\rangle \! \right\rangle_{(0)} \int \frac{\md^3k}{(2\pi)^3}  \left( \frac{H^2}{4 M_{\mathrm{pl}}^2 \epsilon c_s} \right)_H  \frac{1}{p k q}  \nonumber\\
& \frac{3 k}{2 p^4 (k+p+q)^3} \left[s \left(k^3 \left(-33 p^2+30 p q+36 q^2\right) +2 k^4 (9 q-5 p) \right. \right. \nonumber\\
& \left. +k^2 \left(9 p^2 q-42 p^3+34 p q^2+18 q^3 \right) +3 k p (p+q)^2 (p-2 q) \right. \nonumber\\
& \left. +p^2 \left(3 p^2 q-14 p^3+8 p q^2+3 q^3 \right) \right) \nonumber\\
& \left. +k p \left(3 q^2 \left(-4 k^2+9 k p+18 p^2\right) +3 q^3 (7 p-2 k) \right. \right. \nonumber\\
& \left. \left. - p \left(2 k^2 p-3 k^3+33 k p^2+64 p^3\right) +3 q (k+p)^2 (7 p-2 k)\right) \right].
\end{align}
Substituting $q^2 = k^2 - 2 \bp \cdot \bk + p^2$, expanding with respect to the power of $p/k \ll 1$, and extracting terms that have integrand approximately proportional to $k^{-3}$ leads to
\begin{align}
\left\langle \! \left\langle \zeta(\mathbf{p}) \zeta(-\mathbf{p}) \right\rangle \! \right\rangle_{(0,2)}^{[ba]} &= - \frac{1}{16 c_s^2}\left( \frac{2 \tilde{c}_3}{3 c_s^2} \right) \left\langle \! \left\langle \zeta(\mathbf{p}) \zeta(-\mathbf{p}) \right\rangle \! \right\rangle_{(0)}  \nonumber\\
& \hphantom{=} \int \frac{\md^3k}{(2\pi)^3}  \left( \frac{H^2}{4 M_{\mathrm{pl}}^2 \epsilon c_s} \right)_H \left[ \frac{243 s^4}{8 k^3 p^4} - \frac{513 s^2}{16 k^3 p^2} + \frac{45}{16 k^3} \right] \nonumber\\
& = - \frac{1}{32 c_s^2}\left( \frac{2 \tilde{c}_3}{3 c_s^2} \right) \left\langle \! \left\langle \zeta(\mathbf{p}) \zeta(-\mathbf{p}) \right\rangle \! \right\rangle_{(0)} \int_p^\infty \md k ~k^2 \frac{1}{k^3} \left( \frac{H^2}{8 \pi^2 M_{\mathrm{pl}}^2 \epsilon c_s} \right)_H  \nonumber\\
& \hphantom{=} \int_{-1}^1 \md \cos\theta \left[ \frac{243}{8} \cos^4\theta - \frac{513}{16} \cos^2\theta + \frac{45}{16} \right].
\end{align}
Finally, after performing angular integration we get
\begin{equation}
\left\langle \! \left\langle \zeta(\mathbf{p}) \zeta(-\mathbf{p}) \right\rangle \! \right\rangle_{(0,2)}^{[ba]} = \frac{9}{80 c_s^2}\left( \frac{2 \tilde{c}_3}{3 c_s^2} \right) \left\langle \! \left\langle \zeta(\mathbf{p}) \zeta(-\mathbf{p}) \right\rangle \! \right\rangle_{(0)}  \int_p^\infty \md k ~k^2 \frac{1}{k^3} \left( \frac{H^2}{8 \pi^2 M_{\mathrm{pl}}^2 \epsilon c_s} \right)_H .
\end{equation}

\section{Calculation of $\dot{\zeta}(\partial \zeta)^2 \times \dot{\zeta}(\partial \zeta)^2$ term}
In this appendix, we calculate the one-loop correction from $\dot{\zeta}(\partial \zeta)^2 \times \dot{\zeta}(\partial \zeta)^2$ term, that we label as $[bb]$ term. From \eqref{eqbb}, it is given by
\begin{align}
\left\langle \! \left\langle \zeta(\mathbf{p}) \zeta(-\mathbf{p}) \right\rangle \! \right\rangle_{(0,2)}^{[bb]} = & -M_{\mathrm{pl}}^4 \left\langle \! \left\langle \zeta(\mathbf{p}) \zeta(-\mathbf{p}) \right\rangle \! \right\rangle_{(0)} \nonumber\\
& \int_{-\infty}^0 \md \tau_1 \frac{a^2(\tau_1) \epsilon(\tau_1)}{H(\tau_1) c_s^2} \int_{-\infty}^{\tau_1} \md \tau_2 \frac{a^2(\tau_2) \epsilon(\tau_2)}{H(\tau_2) c_s^2}  f(p, \tau_1, \tau_2), \nonumber
\end{align}
\begin{align}
f(p, \tau_1, \tau_2) = \int \frac{\md^3k}{(2\pi)^3}  & \left[ 8(\bp \cdot \bq)^2  \dot{\zeta}_k(\tau_1) \dot{\zeta}^*_k(\tau_2) \zeta_p^*(\tau_1) \zeta_p^*(\tau_2) \zeta_q(\tau_1) \zeta_q^*(\tau_2)  \right. \nonumber\\
& \left. -8 (\bp \cdot \bk) (\bp \cdot \bq) \dot{\zeta}_k(\tau_1) \dot{\zeta}^*_q(\tau_2) \zeta_p^*(\tau_1) \zeta_p^*(\tau_2) \zeta_q(\tau_1) \zeta_k^*(\tau_2) \right. \nonumber\\
& \left.  -8 (\bp \cdot \bq) (\bk \cdot \bq) \dot{\zeta}_k(\tau_1) \dot{\zeta}^*_p(\tau_2) \zeta_p^*(\tau_1) \zeta_k^*(\tau_2) \zeta_q(\tau_1) \zeta_q^*(\tau_2)  \right. \nonumber\\
& \left. -8 (\bp \cdot \bq) (\bk \cdot \bq) \dot{\zeta}^*_p(\tau_1) \dot{\zeta}^*_k(\tau_2) \zeta_k(\tau_1) \zeta_p^*(\tau_2) \zeta_q(\tau_1) \zeta_q^*(\tau_2) \right. \nonumber\\
& \left.  + 4(\bk \cdot \bq)^2  \dot{\zeta}^*_p(\tau_1) \dot{\zeta}^*_p(\tau_2) \zeta_k(\tau_1) \zeta_k^*(\tau_2) \zeta_q(\tau_1) \zeta_q^*(\tau_2) \right]. \label{ef} 
\end{align}
There are five terms in \eqref{ef}, and we will calculate them one by one. The first term is
\begin{align}
\left\langle \! \left\langle \zeta(\mathbf{p}) \zeta(-\mathbf{p}) \right\rangle \! \right\rangle_{(0,2)}^{[bb1]} = & -M_{\mathrm{pl}}^4 \left\langle \! \left\langle \zeta(\mathbf{p}) \zeta(-\mathbf{p}) \right\rangle \! \right\rangle_{(0)} \int_{-\infty}^0 \md \tau_1 \frac{a^2(\tau_1) \epsilon(\tau_1)}{H(\tau_1) c_s^2}  \int_{-\infty}^{\tau_1} \md \tau_2 \frac{a^2(\tau_2) \epsilon(\tau_2)}{H(\tau_2) c_s^2}  \nonumber\\
& \int \frac{\md^3k}{(2\pi)^3} (\bp \cdot \bq)^2 \dot{\zeta}_k(\tau_1) \dot{\zeta}^*_k(\tau_2) \zeta_p^*(\tau_1) \zeta_p^*(\tau_2) \zeta_q(\tau_1) \zeta_q^*(\tau_2). 
\end{align}
Substituting mode function \eqref{modefunction} leads to
\begin{align}
\left\langle \! \left\langle \zeta(\mathbf{p}) \zeta(-\mathbf{p}) \right\rangle \! \right\rangle_{(0,2)}^{[bb1]} = -\left\langle \! \left\langle \zeta(\mathbf{p}) \zeta(-\mathbf{p}) \right\rangle \! \right\rangle_{(0)} & \int \frac{\md^3k}{(2\pi)^3} (p^4 + (\bp \cdot \bk)^2 - 2p^2 \bp \cdot \bk) \nonumber\\
& \left( \frac{H^2}{4 M_{\mathrm{pl}}^2 \epsilon c_s} \right)_H \frac{k}{2 c_s^2 p^3 q^3} f_1(p,q,k),
\end{align}
\begin{align}
f_1(p,q,k) = \int_{-\infty}^0 \md \tau_1 \int_{-\infty}^{\tau_1} \md \tau_2 &~ e^{i c_s (p-k-q) \tau_1} e^{i c_s (p+k+q) \tau_2} \nonumber\\
& (1-i c_s p \tau_1) (1-i c_s p \tau_2) (1 + i c_s q \tau_1) (1 - i c_s q \tau_2).
\end{align}
Then, performing time integration yields
\begin{align}
f_1(p,q,k) = - \frac{1}{4 c_s^2 p^3 (k+p+q)^3} & \left[ -q^2 \left(7 k^2+21 k p+8 p^2\right)+3 p^2 q (5 k+8 p) \right. \nonumber\\
& \left. +p^2 (k+p) (5 k+8 p)-7 q^3 (2 k+3 p)-7 q^4 \right].
\end{align}
Substituting $q^2 = k^2 - 2 \bp \cdot \bk + p^2$, expanding with respect to the power of $p/k \ll 1$, and extracting terms that have integrand approximately proportional to $k^{-3}$ leads to
\begin{align}
\left\langle \! \left\langle \zeta(\mathbf{p}) \zeta(-\mathbf{p}) \right\rangle \! \right\rangle_{(0,2)}^{[bb1]} = & -\frac{1}{2 c_s^4 p^6} \left\langle \! \left\langle \zeta(\mathbf{p}) \zeta(-\mathbf{p}) \right\rangle \! \right\rangle_{(0)}   \nonumber\\
& \int \frac{\md^3k}{(2\pi)^3} [p^4 + (\bp \cdot \bk)^2 - 2p^2 \bp \cdot \bk] \left( \frac{H^2}{4 M_{\mathrm{pl}}^2 \epsilon c_s} \right)_H   \nonumber\\
& \left[ \frac{35 s^2}{16 k^3} - \frac{245s^3}{64k^4} + \frac{441s^4}{164k^5} -\frac{27 p^2}{16k^3} + \frac{355 p^2  s}{64 k^4} + \frac{149 p^4}{64 k^5} - \frac{231 p^2 s^2}{16 k^5} \right].
\end{align}
Next, performing phase-volume integration yields
\begin{align}
\left\langle \! \left\langle \zeta(\mathbf{p}) \zeta(-\mathbf{p}) \right\rangle \! \right\rangle_{(0,2)}^{[bb1]} = & - \frac{1}{2 c_s^4}  \left\langle \! \left\langle \zeta(\mathbf{p}) \zeta(-\mathbf{p}) \right\rangle \! \right\rangle_{(0)} \int \md k ~k^2 \frac{1}{k^3} \left( \frac{H^2}{8 \pi^2 M_{\mathrm{pl}}^2 \epsilon c_s} \right)_H  \nonumber\\
& \frac{2 \pi}{4 \pi} \int_{-1}^1 \md \cos\theta \left[ -\frac{27}{16} + \left( \frac{35}{16} + \frac{149}{64} + 2 \frac{355}{64} \right) \cos^2\theta \right. \nonumber\\
& \left. \hphantom{\frac{2 \pi}{4 \pi} \int_{-1}^1 \md \cos\theta} - \left( \frac{231}{16} + 2\frac{245}{64} \right) \cos^4\theta + \frac{441}{164} \cos^6\theta \right], 
\end{align}
\begin{equation}
\left\langle \! \left\langle \zeta(\mathbf{p}) \zeta(-\mathbf{p}) \right\rangle \! \right\rangle_{(0,2)}^{[bb1]} = - \frac{1}{4 c_s^4} \frac{13}{80} \left\langle \! \left\langle \zeta(\mathbf{p}) \zeta(-\mathbf{p}) \right\rangle \! \right\rangle_{(0)} \int_p^\infty \md k ~k^2 \frac{1}{k^3} \Delta^2_{s(0)}(k_*) \left( \frac{k}{k_*} \right)^{n_s-1}.
\end{equation}
Doing similar calculation for the second term, substituting mode function \eqref{modefunction} leads to
\begin{align}
\left\langle \! \left\langle \zeta(\mathbf{p}) \zeta(-\mathbf{p}) \right\rangle \! \right\rangle_{(0,2)}^{[bb2]} = - \left\langle \! \left\langle \zeta(\mathbf{p}) \zeta(-\mathbf{p}) \right\rangle \! \right\rangle_{(0)} & \int \frac{\md^3k}{(2\pi)^3} [-(\bp \cdot \bk)^2 + p^2 \bp \cdot \bk] \nonumber\\
& \left( \frac{H^2}{4 M_{\mathrm{pl}}^2 \epsilon c_s} \right)_H \frac{1}{2 c_s^2 p^3 k q} f_2(p,q,k),
\end{align}
\begin{align}
f_2(p,q,k) = \int_{-\infty}^0 \md \tau_1 \int_{-\infty}^{\tau_1} \md \tau_2 & ~ e^{i c_s (p-k-q) \tau_1} e^{i c_s (p+k+q) \tau_2} \nonumber\\
& (1-i c_s p \tau_1) (1-i c_s p \tau_2) (1 + i c_s q \tau_1) (1 - i c_s k \tau_2).
\end{align}
Then, performing time integration yields
\begin{align}
f_2(p,q,k) = -\frac{1}{8 c_s^2 p^3 (k+p+q)^3} & \left[ -q^2 \left(28 k^2+51 k p+12 p^2\right)+p (3 k+p) \left(3 k^2+13 k p+16 p^2\right) \right. \nonumber\\
& \left. + q^3 (-(14 k+9 p))+q (p-2 k) (k+p) (7 k+13 p) \right].    
\end{align}
Substituting $q^2 = k^2 - 2 \bp \cdot \bk + p^2$, expanding $p \ll k$, and extracting terms that have integrand approximately proportional to $k^{-3}$ leads to
\begin{align}
\left\langle \! \left\langle \zeta(\mathbf{p}) \zeta(-\mathbf{p}) \right\rangle \! \right\rangle_{(0,2)}^{[bb2]} = & - \frac{1}{2 c_s^4 p^6} \left\langle \! \left\langle \zeta(\mathbf{p}) \zeta(-\mathbf{p}) \right\rangle \! \right\rangle_{(0)} \nonumber\\
& \int \frac{\md^3k}{(2\pi)^3} [-(\bp \cdot \bk)^2 + p^2 \bp \cdot \bk] \left( \frac{H^2}{4 M_{\mathrm{pl}}^2 \epsilon c_s} \right)_H  \nonumber\\
& \left[ \frac{49s^4}{64k^5} + \frac{63p^4}{64k^5} - \frac{27s^2 p^2}{8k^5} - \frac{35s^3}{64k^4} + \frac{127s p^2}{64 k^4}   \right].
\end{align}
Next, performing phase-volume integration yields
\begin{align}
\left\langle \! \left\langle \zeta(\mathbf{p}) \zeta(-\mathbf{p}) \right\rangle \! \right\rangle_{(0,2)}^{[bb2]} = & - \frac{1}{2 c_s^4}  \left\langle \! \left\langle \zeta(\mathbf{p}) \zeta(-\mathbf{p}) \right\rangle \! \right\rangle_{(0)} \int \md k ~k^2 \frac{1}{k^3} \left( \frac{H^2}{8 \pi^2 M_{\mathrm{pl}}^2 \epsilon c_s} \right)_H \nonumber\\
& \frac{2 \pi}{4 \pi}\int_{-1}^1 \md \cos\theta \left[ - \left( \frac{63}{64} + \frac{127}{64} \right) \cos^2\theta + \left( \frac{27}{8} + \frac{35}{64} \right) \cos^4\theta - \frac{49}{64} \cos^6\theta \right], 
\end{align}
\begin{equation}
\left\langle \! \left\langle \zeta(\mathbf{p}) \zeta(-\mathbf{p}) \right\rangle \! \right\rangle_{(0,2)}^{[bb2]} =  \frac{1}{4 c_s^4} \frac{151}{240} \left\langle \! \left\langle \zeta(\mathbf{p}) \zeta(-\mathbf{p}) \right\rangle \! \right\rangle_{(0)} \int_p^\infty \md k ~k^2 \frac{1}{k^3} \Delta^2_{s(0)}(k_*) \left( \frac{k}{k_*} \right)^{n_s-1}.
\end{equation}
Doing similar calculation for the third term, substituting mode function \eqref{modefunction} leads to
\begin{align}
\left\langle \! \left\langle \zeta(\mathbf{p}) \zeta(-\mathbf{p}) \right\rangle \! \right\rangle_{(0,2)}^{[bb3]} = - \left\langle \! \left\langle \zeta(\mathbf{p}) \zeta(-\mathbf{p}) \right\rangle \! \right\rangle_{(0)} & \int \frac{\md^3k}{(2\pi)^3} [ p^2k^2 - (p^2+k^2) \bp \cdot \bk + (\bp \cdot \bk)^2 ] \nonumber\\
& \left( \frac{H^2}{4 M_{\mathrm{pl}}^2 \epsilon c_s} \right)_H \frac{1}{2 c_s^2 p k q^3} f_3(p,q,k),
\end{align}
\begin{align}
f_3(p,q,k) = \int_{-\infty}^0 \md \tau_1 \int_{-\infty}^{\tau_1} \md \tau_2 & ~ e^{i c_s (p-k-q) \tau_1} e^{i c_s (p+k+q) \tau_2} \nonumber\\
& (1-i c_s p \tau_1) (1-i c_s q \tau_2) (1 + i c_s q \tau_1) (1 - i c_s k \tau_2).
\end{align}
Then, performing time integration yields
\begin{align}
f_3(p,q,k) = -\frac{1}{8 c_s^2 p^4 (k+p+q)^3} & \left[ q^4 (-(9 k+5 p)) +3 q^2 \left(-11 k^2 p-3 k^3-11 k p^2+p^3\right) \right. \nonumber\\
& \left. + 6 p^2 q \left(2 k^2+8 k p+3 p^2\right)+2 p^2 (k+p) \left(2 k^2+8 k p+3 p^2\right) \right. \nonumber\\
& \left. -2 q^3 \left(9 k^2+19 k p+7 p^2\right) \right]
\end{align}
Substituting $q^2 = k^2 - 2 \bp \cdot \bk + p^2$, expanding with respect to the power of $p/k \ll 1$, and extracting terms that have integrand approximately proportional to $k^{-3}$ leads to
\begin{align}
\left\langle \! \left\langle \zeta(\mathbf{p}) \zeta(-\mathbf{p}) \right\rangle \! \right\rangle_{(0,2)}^{[bb3]} = & - \frac{1}{2 c_s^4 p^6}  \left\langle \! \left\langle \zeta(\mathbf{p}) \zeta(-\mathbf{p}) \right\rangle \! \right\rangle_{(0)} \nonumber\\
& \int \frac{\md^3k}{(2\pi)^3} [ p^2k^2 - (p^2+k^2) \bp \cdot \bk + (\bp \cdot \bk)^2] \left( \frac{H^2}{4 M_{\mathrm{pl}}^2 \epsilon c_s} \right)_H  \nonumber\\
& \left[ -\frac{17sp}{32k^4} + \frac{115s^2p}{128k^5} - \frac{203s^3}{128k^6} - \frac{131p^3}{128k^5} + \frac{439sp^3}{128k^6}   \right].
\end{align}
Next, performing phase-volume integration yields
\begin{align}
\left\langle \! \left\langle \zeta(\mathbf{p}) \zeta(-\mathbf{p}) \right\rangle \! \right\rangle_{(0,2)}^{[bb3]} = & - \frac{1}{2 c_s^4}  \left\langle \! \left\langle \zeta(\mathbf{p}) \zeta(-\mathbf{p}) \right\rangle \! \right\rangle_{(0)} \int \md k ~k^2 \frac{1}{k^3} \left( \frac{H^2}{8 \pi^2 M_{\mathrm{pl}}^2 \epsilon c_s} \right)_H \nonumber\\
& \frac{2 \pi}{4 \pi}\int_{-1}^1 \md \cos\theta \left[ - \frac{131}{128} + \left( \frac{115}{128} - \frac{17}{32} - \frac{131}{128} + \frac{439}{128} \right) \cos^2\theta \right. \nonumber\\
& \left. \hphantom{\frac{2 \pi}{4 \pi}\int_{-1}^1 \md \cos\theta} + \left( \frac{115}{128} - \frac{203}{128} \right) \cos^4\theta \right] 
\end{align}
\begin{equation}
\left\langle \! \left\langle \zeta(\mathbf{p}) \zeta(-\mathbf{p}) \right\rangle \! \right\rangle_{(0,2)}^{[bb3]} = \frac{1}{4 c_s^4} \frac{227}{480} \left\langle \! \left\langle \zeta(\mathbf{p}) \zeta(-\mathbf{p}) \right\rangle \! \right\rangle_{(0)} \int_p^\infty \md k ~k^2 \frac{1}{k^3} \Delta^2_{s(0)}(k_*) \left( \frac{k}{k_*} \right)^{n_s-1}.
\end{equation}
Doing similar calculation for the fourth term, substituting mode function \eqref{modefunction} leads to
\begin{align}
\left\langle \! \left\langle \zeta(\mathbf{p}) \zeta(-\mathbf{p}) \right\rangle \! \right\rangle_{(0,2)}^{[bb4]} = - \left\langle \! \left\langle \zeta(\mathbf{p}) \zeta(-\mathbf{p}) \right\rangle \! \right\rangle_{(0)} & \int \frac{\md^3k}{(2\pi)^3} [ p^2k^2 - (p^2+k^2) \bp \cdot \bk + (\bp \cdot \bk)^2 ] \nonumber\\
& \left( \frac{H^2}{4 M_{\mathrm{pl}}^2 \epsilon c_s} \right)_H \frac{1}{2 c_s^2 p k q^3} f_4(p,q,k),
\end{align}
\begin{align}
f_4(p,q,k) = \int_{-\infty}^0 \md \tau_1 \int_{-\infty}^{\tau_1} \md \tau_2 &~ e^{i c_s (p-k-q) \tau_1} e^{i c_s (p+k+q) \tau_2} \nonumber\\
& (1 + i c_s k \tau_1) (1-i c_s q \tau_2) (1 + i c_s q \tau_1) (1 - i c_s p \tau_2).
\end{align}
Then, performing time integration yields
\begin{align}
f_4(p,q,k) = \frac{1}{8 c_s^2 p^4 (k+p+q)^3} & \left[ -2 q^3 \left(9 k^2+8 k p-8 p^2\right)-3 q^2 (3 k+p) \left(k^2+2 k p-p^2\right) + \right. \nonumber\\
& \left. 6 p^2 q \left(2 k^2-5 p^2\right)-2 p^2 (k+p) \left(5 p^2-2 k^2\right)+q^4 (5 p-9 k) \right].
\end{align}
\begin{align}
\left\langle \! \left\langle \zeta(\mathbf{p}) \zeta(-\mathbf{p}) \right\rangle \! \right\rangle_{(0,2)}^{[bb4]} = & - \frac{1}{2 c_s^4 p^6} \left\langle \! \left\langle \zeta(\mathbf{p}) \zeta(-\mathbf{p}) \right\rangle \! \right\rangle_{(0)} \nonumber\\
& \int \frac{\md^3k}{(2\pi)^3} [ p^2k^2 - (p^2+k^2) \bp \cdot \bk + (\bp \cdot \bk)^2 ] \left( \frac{H^2}{4 M_{\mathrm{pl}}^2 \epsilon c_s} \right)_H   \nonumber\\
& \left[ -\frac{17sp}{32k^4} + \frac{115s^2p}{128k^5} - \frac{203s^3}{128k^6} - \frac{131p^3}{128k^5} + \frac{439sp^3}{128k^6}   \right]. 
\end{align}
The fourth term has the same contribution as the third term. Doing similar calculation for the fifth term, substituting mode function \eqref{modefunction} leads to
\begin{align}
\left\langle \! \left\langle \zeta(\mathbf{p}) \zeta(-\mathbf{p}) \right\rangle \! \right\rangle_{(0,2)}^{[bb5]} = - \left\langle \! \left\langle \zeta(\mathbf{p}) \zeta(-\mathbf{p}) \right\rangle \! \right\rangle_{(0)} & \int \frac{\md^3k}{(2\pi)^3} [ k^4 + (\bp \cdot \bk)^2 - 2 k^2 \bp \cdot \bk ] \nonumber\\
& \left( \frac{H^2}{4 M_{\mathrm{pl}}^2 \epsilon c_s} \right)_H \frac{p}{4 c_s^2 k^3 q^3} f_5(p,q,k),
\end{align}
\begin{align}
f_5(p,q,k) = \int_{-\infty}^0 \md \tau_1 \int_{-\infty}^{\tau_1} \md \tau_2 & ~ e^{i c_s (p-k-q) \tau_1} e^{i c_s (p+k+q) \tau_2} \nonumber\\
& (1 + i c_s q \tau_1) (1-i c_s q \tau_2) (1 + i c_s k \tau_1) (1 - i c_s k \tau_2).
\end{align}
Then, performing time integration yields
\begin{align}
f_5(p,q,k) = -\frac{1}{4 c_s^2 p^5 (k+p+q)^3} & \left[ q^4 \left(3 k^2-p^2\right)+3 q^3 (2 k+p) \left(k^2+k p-p^2\right) \right. \nonumber\\
& \left. +q^2 \left(4 k^2 p^2+9 k^3 p+3 k^4-6 k p^3+2 p^4\right) \right. \nonumber\\
& \left. - 3 p^2 q \left(2 k^2 p+k^3-4 k p^2-2 p^3\right) \right. \nonumber\\
& \left. +p^2 (k+p) \left(-2 k^2 p-k^3+4 k p^2+2 p^3\right) \right]. 
\end{align}
Substituting $q^2 = k^2 - 2 \bp \cdot \bk + p^2$, expanding with respect to the power of $p/k \ll 1$, and extracting terms that have integrand approximately proportional to $k^{-3}$ leads to
\begin{align}
\left\langle \! \left\langle \zeta(\mathbf{p}) \zeta(-\mathbf{p}) \right\rangle \! \right\rangle_{(0,2)}^{[bb5]} = & - \frac{1}{4 c_s^4 p^6} \left\langle \! \left\langle \zeta(\mathbf{p}) \zeta(-\mathbf{p}) \right\rangle \! \right\rangle_{(0)} \nonumber\\
& \int \frac{\md^3k}{(2\pi)^3} [ k^4 + (\bp \cdot \bk)^2 - 2 k^2 \bp \cdot \bk ] \left( \frac{H^2}{4 M_{\mathrm{pl}}^2 \epsilon c_s} \right)_H   \nonumber\\
& \left[ - \frac{15 s^2}{16 k^5} + \frac{105s^3}{64k^6} - \frac{189s^4}{64k^7} + \frac{11p^2}{16k^5} - \frac{125sp^2}{64k^6} + \frac{77s^2p^2}{16k^7} - \frac{99p^4}{64k^7}    \right].
\end{align}
Next, performing phase-volume integration yields
\begin{align}
\left\langle \! \left\langle \zeta(\mathbf{p}) \zeta(-\mathbf{p}) \right\rangle \! \right\rangle_{(0,2)}^{[bb5]} = & - \frac{1}{4 c_s^4}  \left\langle \! \left\langle \zeta(\mathbf{p}) \zeta(-\mathbf{p}) \right\rangle \! \right\rangle_{(0)} \int \md k ~k^2 \frac{1}{k^3} \left( \frac{H^2}{8 \pi^2 M_{\mathrm{pl}}^2 \epsilon c_s} \right)_H \nonumber\\
& \frac{2 \pi}{4 \pi}\int_{-1}^1 \md \cos\theta \left[ - \frac{99}{64} + \left( \frac{11}{16} + \frac{77}{16} - 2 \frac{125}{64} \right) \cos^2\theta \right. \nonumber\\
& \left. \hphantom{\frac{2 \pi}{4 \pi}\int_{-1}^1 \md \cos\theta} + \left( - \frac{15}{16} + 2\frac{105}{64} - \frac{189}{64} \right)  \cos^4\theta \right], 
\end{align}
\begin{equation}
\left\langle \! \left\langle \zeta(\mathbf{p}) \zeta(-\mathbf{p}) \right\rangle \! \right\rangle_{(0,2)}^{[bb5]} = \frac{1}{8 c_s^4} \frac{91}{40} \left\langle \! \left\langle \zeta(\mathbf{p}) \zeta(-\mathbf{p}) \right\rangle \! \right\rangle_{(0)} \int_p^\infty \md k ~k^2 \frac{1}{k^3} \Delta^2_{s(0)}(k_*) \left( \frac{k}{k_*} \right)^{n_s-1}.
\end{equation}
Finally, after adding all terms we get
\begin{equation}
\left\langle \! \left\langle \zeta(\mathbf{p}) \zeta(-\mathbf{p}) \right\rangle \! \right\rangle_{(0,2)}^{[bb]} = \frac{1}{4 c_s^4} \frac{51}{20} \left\langle \! \left\langle \zeta(\mathbf{p}) \zeta(-\mathbf{p}) \right\rangle \! \right\rangle_{(0)} \int_p^\infty \md k ~k^2 \frac{1}{k^3} \Delta^2_{s(0)}(k_*) \left( \frac{k}{k_*} \right)^{n_s-1}.
\end{equation}

\acknowledgments

J. K. acknowledges the support from JSPS KAKENHI Grant No.~22J20289 and Global Science Graduate Course (GSGC) program of The University of Tokyo. J. Y. is supported by JSPS KAKENHI Grant No.~20H05639 and Grant-in-Aid for Scientific Research on Innovative Areas 20H05248. 

\bibliographystyle{jhep}
\bibliography{Reference}

\end{document}